\documentclass[]{aa} 
\usepackage{natbib}
\usepackage{amsmath}
\usepackage{graphicx}
\usepackage{txfonts}   
\usepackage{lscape}
\usepackage{gensymb}
\usepackage{color}
\usepackage{newtxtext,newtxmath}
\usepackage{pdflscape}
\usepackage[hyperindex,breaklinks=true, colorlinks, citecolor=blue]{hyperref}
\usepackage[switch,pagewise]{lineno}

\DeclareRobustCommand{\ion}[2]{\textup{#1\,\textsc{\lowercase{#2}}}}

\DeclareRobustCommand{\kms}{km\,${\rm s}^{-1}$}
\DeclareRobustCommand{\jyb}{Jy\,beam${}^{-1}$\,}

\title{Faint absorption of the ground state hyperfine-splitting transitions of hydroxyl at 18 cm in the Galactic Disk\thanks{This work is based on observations made with the Karl G. Jansky Very Large Array under the project VLA 17A-055.  An earlier version of parts of this work appeared in the PhD thesis of M. R. Rugel.}}

  \author{
  	M. R. Rugel\inst{1,2,3,}\thanks{M. R. Rugel is a Jansky Fellow of the National Radio Astronomy Observatory}\and
	H. Beuther\inst{4} \and
	J. D. Soler\inst{5} \and
	P. Goldsmith\inst{6}\and
	L. Anderson\inst{7,8,9} \and
	A. Hafner\inst{10}\and
	J. R. Dawson\inst{11,12}\and
	Y. Wang\inst{4}\and
	S. Bihr\inst{4}\and
	H. Wiesemeyer\inst{3}\and
	R. G\"usten\inst{3}\and
	M.-Y. Lee\inst{13,14} \and
	D. Riquelme\inst{3,15}\and
	A.~M. Jacob\inst{3,16}\and
	W.-J. Kim\inst{16}\and
	M. Busch\inst{17}\and
	S. Khan\inst{3}\and
	A. Brunthaler\inst{3}
	}
  \institute{
    National Radio Astronomy Observatory, PO Box O, 1003 Lopezville Road, Socorro, NM 87801, USA \and
    Center for Astrophysics | Harvard \& Smithsonian, 60 Garden Street, Cambridge, MA 02138, USA \and
    Max Planck Institute for Radio Astronomy, Auf dem H\"ugel 69, 53121 Bonn, Germany \and
    Max Planck Institute for Astronomy, K\"onigstuhl 17, 69117 Heidelberg, Germany \and
    Istituto di Astrofisica e Planetologia Spaziali (IAPS). INAF. Via Fosso del Cavaliere 100, 00133 Roma, Italy\and    
    Jet Propulsion Laboratory, California Institute of Technology, 4800 Oak Grove Drive, Pasadena, 91109, USA \and
    Department of Physics and Astronomy, West Virginia University, Morgantown, WV 26506, USA \and
    Center for Gravitational Waves and Cosmology, West Virginia University, Chestnut Ridge Research Building, Morgantown, WV 26505, USA\and
    Adjunct Astronomer at the Green Bank Observatory, PO Box 2, Green Bank, WV 24944, USA\and
    School of Physics, The University of Sydney, Physics Rd, Camperdown NSW 2050, Australia\and
    School of Mathematical and Physical Sciences and Astrophysics and Space Technologies Research Centre, Macquarie University, 2109, NSW, Australia \and
    Australia Telescope National Facility, CSIRO Space \& Astronomy, PO Box 76, Epping, NSW 1710, Australia \and
    Korea Astronomy and Space Science Institute, 776 Daedeok-daero, Daejeon 34055, Republic of Korea\and
    Department of Astronomy and Space Science, University of Science and Technology, 217 Gajeong-ro, Daejeon 34113, Republic of Korea\and
    Departamento de Astronomía, Universidad de la Serena, Raúl Bitrán 1305, la Serena, Chile\and
    I. Physikalisches Institut, Universit\"at zu K\"oln, Z\"ulpicher Str. 77, 50937 K\"oln, Germany\and
    Department of Astronomy \& Astrophysics, University of California, San Diego, 9500 Gilman Drive, San Diego, CA 92093, USA
	}
	
 \abstract
 {The interstellar hydride hydroxyl (OH) is a potential tracer of CO-dark molecular gas. We present new high-sensitivity absorption line observations of the four ground state hyperfine-splitting transitions of OH at 18-cm wavelength towards four Galactic and extra-galactic continuum sources as follow-up to the THOR survey. We compare these to deep observations of the [\ion{C}{ii}] 158\,$\mu$m line at 1.9 THz obtained with the upGREAT instrument on SOFIA, observations of the neutral atomic hydrogen (\ion{H}{i}) 21 cm line with the VLA, and CO ($J$ = 2--1) lines obtained with the APEX PI230 receiver at the APEX 12\,m sub-mm telescope. We trace OH over a large range of molecular hydrogen column densities of 7.9$\times10^{19}$\,cm$^{-2}$ to 4.7$\times10^{22}$\,cm$^{-2}$, and derive OH abundances with respect to molecular and total hydrogen column densities of X$_{\rm OH, H_2} =N_{\rm OH}/N_{\rm H_2} = 1.2^{+0.3}_{-0.2}\times10^{-7}$  and $X_{\rm OH, H} = N_{\rm OH}/N_{\rm H}=4.8^{+0.9}_{-0.8}\times10^{-8}$ , respectively. Increased sensitivity and spectral resolution allowed us to detect weak and narrow features with the lowest column density detected at $N_{\rm OH}= 3.7\times10^{13}$\,cm$^{-2}$. The increase in sensitivity is a factor of 5 in direct comparison at the resolution the OH observations in the THOR survey (1.5\,\kms). We identify only one OH absorption component out of 23 without CO counterpart, yet several with intermediate molecular gas fractions (${\rm f_{mol}\leq0.8}$). A potential association of [\ion{C}{ii}] 158\,$\mu$m emission with an OH absorption component is seen toward one sightline. Our results confirm that OH absorption traces molecular gas across diffuse and dense environments of the interstellar medium. At the sensitivity limits of the present observations our detection of only one CO-dark molecular gas feature appears in agreement with previous studies. We conclude that if OH absorption was to be used as a CO-dark molecular gas tracer, deeper observations or stronger background targets are necessary to unveil its full potential as a CO-dark molecular gas tracer, and yet it will never be an exclusive tracer of CO-dark molecular gas. For OH hyperfine-splitting transitions in the vicinity of photodissociation regions in W43-South, we detect a spectral and spatial offset between the peak of the inversion of the OH 1612 MHz line and the absorption of the OH 1720 MHz line on the one hand, and the absorption of the OH main lines on the other hand, which provides additional constraints on the interpretation of the OH 18\,cm line signatures typical of HII regions.
} 

\date{Received XXX; accepted XXX}

\begin{document} 
\maketitle

\section{Introduction}

\begin{figure*}[!ht]
\centering
  \includegraphics[width=0.99\textwidth]{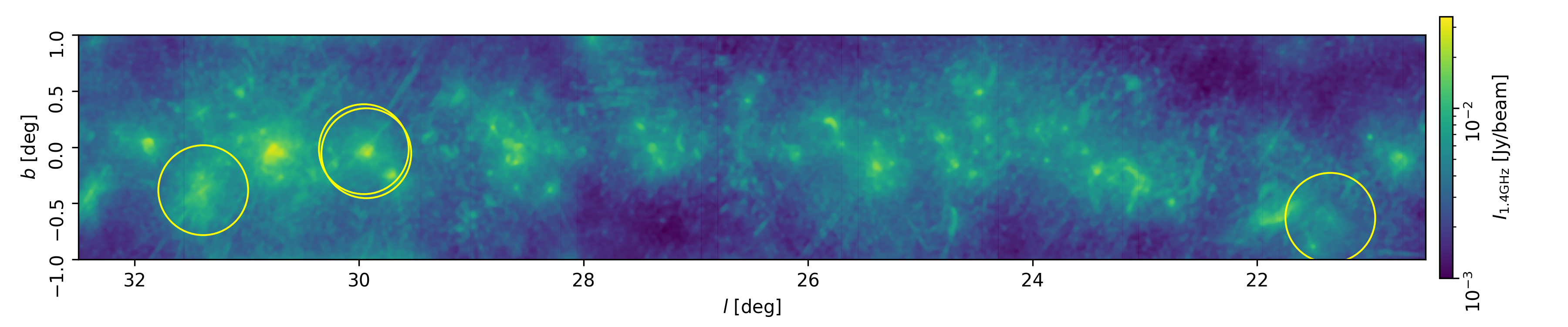}
  \includegraphics[width=0.99\textwidth]{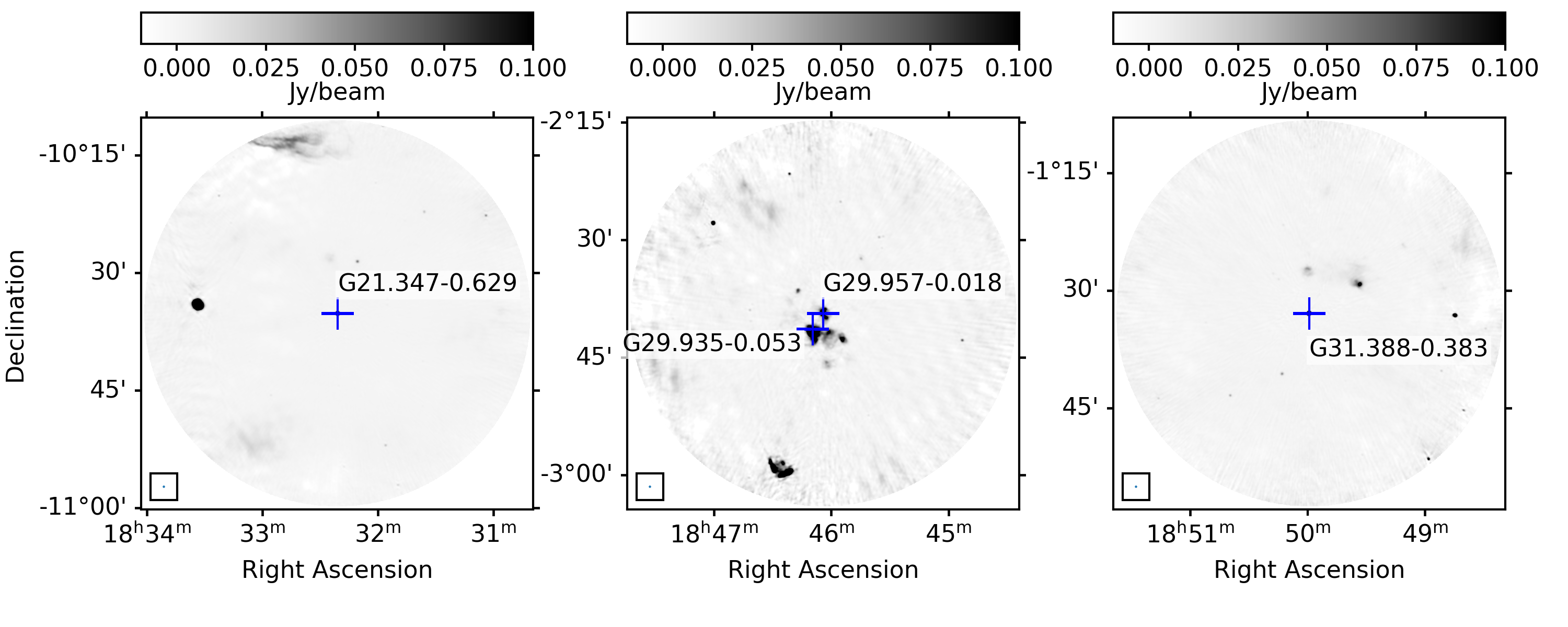}
  \caption{{\it Top:} THOR 1.4\,GHz continuum emission in the Galactic coordinate frame \citep{WangBeuther:2020aa}. The sources discussed here are located at the center of the yellow circles. {\it Bottom:} VLA 1.4 GHz continuum towards the four positions analyzed in this work {\it (blue crosses)} in equatorial coordinates. The images are smoothed to an angular resolution of 18\arcsec.}
  \label{fig:sample_selection}
\end{figure*}

Molecular clouds are the birthplaces of stars. They form out of the diffuse atomic interstellar medium (ISM) once sufficient shielding from the interstellar radiation field is provided \citep[e.g.,][and references therein]{DobbsKrumholz:2014aa,ChevanceKrumholz:2023vt,PinedaArzoumanian:2023ab}, which also enables star formation \citep{GloverClark:2012aa}. It is essential to study the formation of molecular clouds for our understanding of star formation in the Milky Way. 

Gas in cold molecular clouds consists mostly of H$_2$. H$_2$ emission from molecular clouds can be detected in shocks \citep[e.g.,][see also discussion in \citealt{Sternberg:1989jf}]{BeckwithEvans:1983ur,Beuthervan-Dishoeck:2023sf} or by fluorescence emission \citep[e.g.,][]{Sellgren:1986yz,WittStecher:1989je,Sternberg:1989dx,BurkhartDharmawardena:2025mr}. Its rotational transitions with excitation energies about 500\,K are impossible to be observed in emission under cold ISM conditions. Alternatively, such clouds are commonly studied with observations of the CO molecule \citep{WilsonJefferts:1970of}, whose low angular momentum ($J$) rotational lines are easily accessible at millimeter wavelengths. Several studies have indicated the presence of molecular gas not traced by CO emission \citep[e.g.,][]{GrenierCasandjian:2005aa,WolfireHollenbach:2010aa,SmithGlover:2014aa,Planck-CollaborationAde:2011aa}. CO-dark molecular hydrogen gas, here defined as gas without detection of $^{12}$CO emission, has been found across the Galaxy at significant fractions \citep{PinedaLanger:2013aa,LangerVelusamy:2014aa,BuschEngelke:2021nk}, in particular in regions of low extinction \citep{ParadisDobashi:2012aa,XuLi:2016aa,RemyGrenier:2018ab}. Observationally constraining the properties of ``CO-dark'' gas or simply ``dark'' gas thus contributes to our understanding of both the \ion{H}{i}-to-H$_2$ conversion, as well as the onset of molecular cloud and star formation.

The ground state hyperfine-splitting (HFS) transitions of the hydroxyl molecule (OH) at cm wavelengths have been detected in regions with CO-dark gas. \citet{XuLi:2016aa} have investigated the relation between OH emission and dark gas fractions across the molecular cloud boundary in Taurus. \citet{AllenHogg:2015aa}, \citet{BuschAllen:2019cr} and \citet{BuschEngelke:2021nk} detected OH emission towards CO-dark lines-of-sight in the outer Galaxy, and \citet{LiTang:2018ab} found OH absorption components without CO counterparts towards strong extragalactic continuum sources at high Galactic latitudes. Other detailed Galactic plane surveys investigating OH in emission across the Galactic plane include The Southern Parkes Large-Area Survey in Hydroxyl (SPLASH; \citealt{DawsonJones:2022kj}). OH has also been found in high-mass star forming regions in association with photon-dominated/photo-dissociation regions (PDRs) and outflows \citep[e.g.,][]{GoicoecheaJoblin:2011gi,WampflerBruderer:2011we,WampflerBruderer:2013aa}. 
The abundance of OH has been analyzed in several studies using the 18\,cm lines \citep[e.g.,][]{Goss:1968aa,Crutcher:1979aa,LisztLucas:1996aa,LisztLucas:2002aa}, with a detailed, recent studies of OH absorption presented in \citet{HafnerDawson:2023cg} and \citet{NguyenDawson:2018xt}. The OH abundance was also investigated with the fundamental OH rotational transitions into the ground state at far-infrared wavelengths \citep{WiesemeyerGusten:2016aa,JacobNeufeld:2022aa}, as well as through its electronic transition at ultra-violett wavelengths \citep[e.g.][]{WeselakGalazutdinov:2010aa}. 
Given that the OH hyperfine splitting transitions at 18-cm are readily accessible at ever higher resolution and sensitivity by today's and future radio telescopes, a detailed study of OH in different Galactic environments seems warranted. 

In this work, we present deep OH absorption observations along the lines of sight towards three bright background sources, as a follow-up study to The \ion{H}{i}, OH, Recombination Line survey (THOR;  \citealt{BeutherBihr:2016aa,WangBeuther:2020aa}), which observed the OH 18 cm lines over a large fraction of the first Galactic quadrant \citep{RugelBeuther:2018aa,BeutherWalsh:2019ng}. We discuss the OH absorption towards four continuum sources which are chosen to be strong enough to reveal faint OH absorption from diffuse sightline clouds with typical optical depths of $\tau<$0.1 \citep[e.g.,][]{DickeyCrovisier:1981aa,LisztLucas:1996aa}, and increased spectral resolution as compared to THOR to resolve narrow absorption features. The field of view of one of the targets includes an extended \ion{H}{ii} region complex, which enables additional studies of OH in PDRs \citep[e.g.,][]{BroganTroland:2001kf} towards \ion{H}{ii} regions. 

Furthermore, we search for associated [\ion{C}{ii}] emission using observations carried out with the Stratospheric Observatory for Far-Infrared  Astronomy (SOFIA; \citealt{YoungBecklin:2012lq}). [\ion{C}{ii}] emission can originate from different ISM environments, including the transition region between atomic and molecular gas. It has been used to estimate the CO-dark gas fraction in individual clouds \citep{XuLi:2016aa}, as well over the entire Milky Way disk \citep{PinedaLanger:2013aa}. [\ion{C}{ii}] emission has been directly detected in the diffuse ISM \citep{GoldsmithPineda:2018aa}, and may trace gas accretion onto molecular clouds \citep{HeyerGoldsmith:2022xg}. [\ion{C}{ii}] is readily detected in absorption along the line-of-sight in clouds of varying molecular gas fractions \citep[e.g.,][]{GerinRuaud:2015aa,JacobNeufeld:2022aa}. 

The goal of this work is to characterize the abundance of the OH molecule in different ISM conditions. Of particular interest are the properties of the OH gas in diffuse ISM environments where CO becomes a less-reliable tracer of the bulk molecular hydrogen content. We will present comparisons between [\ion{C}{ii}] emission and OH absorption. We additionally report findings on the excitation conditions of OH in \ion{H}{ii} regions.

\begin{table*}
\centering
\caption{Target locations and 1.4\,GHz continuum fluxes}
\label{tbl:observations}
\begin{tabular}{lccccc}
\hline
\hline
Source  & $l$ & $b$ & R. A. (J2000) & Dec. (J2000)  & $F_{\rm cont, peak}$(1.4\,GHz at 18$^{\prime\prime}$) \\
& $^\circ$&  $^\circ$ & (${\rm ^h\,^m\,^s}$) & (${\rm ^\circ\,^m\,^s}$) & [\jyb] \\
\hline
G21.347$-$0.629 & 21.347 & $-$0.629 & 18 32 20.9 & $-$10 35 11.4  &0.9 \\
G29.957$-$0.018 & 29.957 & $-$0.018 & 18 46 04.2 & $-$02 39 19.3  &1.0 \\
G29.935$-$0.053 & 29.935 & $-$0.054 &18 46 09.6 & $-$02 41 27.5 & 0.4 \\
G31.388$-$0.383 & 31.388 & $-$0.383 & 18 49 59.2 & $-$01 32 56.0 & 1.3 \\
\hline
\end{tabular}
\end{table*}

\section{Observations}\label{sec:obs}

\subsection{VLA observations}
We selected four continuum sources with a flux of $\sim$1\,\jyb (Table~\ref{tbl:observations}). The \ion{H}{ii} region G29.957$-$0.018 has a spectral index\footnote{The spectral index is defined as $S_\nu = S_0 \times \left(\frac{\nu}{\nu_0}\right)^\alpha$, where $S_0$ is the flux at a reference frequency, $\nu_0$ at 1.4\,GHz} of $\alpha=0.9$ \citep{WangBihr:2018aa}, where positive spectral indices indicate thermal emission. G29.957$-$0.018 and G29.935$-$0.053 are part of the \ion{H}{ii} region complex W43-South, which has been attributed to the W43 complex \citep{MotteNguyen-Luong:2014aa}. The continuum emission in W43-South at cm wavelengths has presented in \citet{BeltranOlmi:2013aa}, and the entire W43 complex is located at a distance of 5.49\,kpc \citep{ZhangMoscadelli:2014aa}. Two sources are of extragalactic origin (G21.347$-$0.629 and G31.388$-$0.383; with $\alpha=0.0$ and $\alpha=-0.9$). All sources were selected based on OH absorption components in the THOR survey \citep{RugelBeuther:2018aa} with weak (G29.957$-$0.018) or no (G31.388$-$0.383) $^{13}$CO (1--0) emission counterparts in the Galactic Ring Survey (GRS; \citealt{JacksonRathborne:2006aa}).

We chose a channel width of 0.488\,kHz (0.1\,\kms\ at 1420\,MHz), which is a significant improvement in comparison to the THOR survey (1.5\,\kms; \citealt{RugelBeuther:2018aa}).
The bandwidth is 4~MHz for the \ion{H}{I} and OH 1665/1667 MHz bands, and 2 MHz bands for the OH 1612 and OH 1720 MHz lines, which results in at least 300\,\kms\ of usable bandwidth for the OH transitions (see Table~\ref{tbl:frequencies} for frequencies discussed in this work).
We observed for 8 nights between June and August 2017 with the VLA in C-configuration, with a total integration time of 7\,h for G21.347$-$0.629, 4\,h for G29.957$-$0.018 and G29.935$-$0.053, and 3\,h for G31.388$-$0.383. The observations were split into blocks of 2 hours. We observed 3C286 for 10-20 minutes for bandpass calibration and used J1822-0938 as a complex gain calibrator. 

We processed the observational data with the CASA\footnote{\url{http://casa.nrao.edu}; version 4.7.2; pipeline version 1.3.11} calibration pipeline. We inspected the data for antenna and baseline artifacts. The calibrated data were imaged and de-convolved using the CASA task {\tt clean}. We subtracted the continuum from line-free channels before imaging, and imaged spectral line cubes and continuum emission separately. The restoring beam is between 17\farcs2$\times$12\farcs8 and 12\farcs8$\times$11\farcs5. We smoothed all data to 18\arcsec\ resolution (full width at half maximum, FWHM). The noise in a 0.2 \kms-wide channel is between 5--7\,m\jyb\ for the OH transitions. The increase in sensitivity is a factor of 5 in direct comparison at the resolution the OH observations in the THOR survey (10\,m\jyb\ at 1.5\,\kms\ spectral resolution). For the \ion{H}{i} line, the sensitivity is around 5\,m\jyb\ in a 0.5 \kms-wide channel, corresponding to a brightness temperature of $\sim$10\,K. At velocities of Galactic \ion{H}{I} emission, the noise increases up to a factor of two due to significant contributions of the \ion{H}{I} emission to the system temperature.

\subsection{SOFIA [CII] observations}
We used the upGREAT receiver on the SOFIA telescope \citep{RisacherGusten:2016yi} to measure the [\ion{C}{ii}] 158\,$\mu$m line (see Table~\ref{tbl:frequencies}) towards G29.935$-$0.053 and G31.388$-$0.383. Due to observing time constraints, we were unable to obtain observations towards G21.347$-$0.629 and G29.957$-$0.018 was restricted to guaranteed time programs at the time of the proposal. The observations were conducted on June 8, 13, 19 and 20 of 2018 during flights from Christchurch, New Zealand, at an altitude of about 13 km as part of the project 06\_0171 in position-switching total power mode. We used the standard SOFIA GREAT data reduction pipeline kalibrate. The spectra were corrected for atmospheric absorption. 
 
The SOFIA upGREAT instrument consists of an array of 7 receivers (``pixels'') which observe simultaneously and and comprises two frequency channels operating at 1.9\,THz and 4.7\,THz, respectively. The low-frequency array was tuned to 1.9 THz to cover the [\ion{C}{ii}] 158\,$\mu$m line, with an angular resolution of 14\farcs1. The high-frequency array was tuned to 4.7 THz to cover the [\ion{O}{i}] 63\,$\mu$m line (see Table~\ref{tbl:frequencies}), with an angular resolution of 6\farcs3. The data were calibrated with a forward efficiency of $\eta_{\rm f}$=0.97 and a main-beam efficiency which varies among the pixels between 0.55 and 0.70. As [\ion{C}{ii}] 158\,$\mu$m emission extends to high Galactic latitudes, we chose two separate reference positions for both sources, identified as regions with low CO emission and low dust opacity using data from the Planck mission \citep{Planck-CollaborationAde:2011aa,Planck-CollaborationAbergel:2014wl} in absence of large scale maps of [\ion{C}{ii}] 158\,$\mu$m emission. The time on each source was equally shared between the on-target position (50\,\%) and two reference positions (25\,\% each). 

Pixel 2 was only available in vertical polarization, while all other pixels were observed in both linear polarizations. We note that pixel 4 shows higher baseline instabilities.
Overall baselines are subtracted with first-order polynomials. For each pixel, we averaged all polarizations and observations with different reference positions. Similar steps have been applied for the [\ion{O}{i}] 63\,$\mu$m line, for which only the vertical polarization orientation was available. In the end, we averaged all available polarizations and spectra obtained with different reference positions. The final RMS on main beam temperature scales in the central pixel is $\sigma(T_{\rm mb})\sim$0.1\,K in a 0.39\,\kms\ channel for the [\ion{C}{ii}] 158\,$\mu$m, and $\sim$0.2\,K in a 0.15\,\kms\ channel for the [\ion{O}{i}] 63\,$\mu$m spectra. Residual baseline ripples are present in both lines at a low level. We flag any remaining channels with potential contamination from emission in the reference position, and disregard them in the following analysis. All SOFIA upGREAT spectra are shown in Appendix~\ref{sec:app_sofia}. 

For G29.935$-$0.053, we searched for the [${}^{13}$\ion{C}{ii}] line. While we see indications of the $F$=1--0 line, the $F$=0--1 line is blended with the [\ion{C}{ii}] 158\,$\mu$m line, and the $F$=1--1 line remains undetected. The velocity of the faint [${}^{13}$\ion{C}{ii}] $F$=1--0 line aligns with the velocity of a potential [\ion{C}{ii}] 158\,$\mu$m self-absorption dip in many of the pixels. Due to overlap in velocity with the complex [\ion{C}{ii}] 158\,$\mu$m emission along the line of sight across the Galaxy, a clear identification of the [${}^{13}$\ion{C}{ii}] $F$=1--0 remains difficult.

\subsection{CO observations with the APEX telescope and from archival data}\label{sec:obs_archival_co}

We observed the $J=$2--1 lines of different isotopologues of the CO molecule with the PI230 instrument on the Atacama Pathfinder EXperiment (APEX) 12\,m submillimeter telescope (\citealt{GustenNyman:2006uw}) for the sources for which we had obtained SOFIA [\ion{C}{ii}] 158\,$\mu$m observations, G31.388$-$0.383 and G29.935$-$0.053.
We observed both sources in position-switching mode. The tuning frequencies were chosen to cover the $^{12}$CO, $^{13}$CO, and C$^{18}$O isotopologues, as well as the H30$\alpha$ radio recombination line. The CO lines have been resampled to a spectral resolution of 0.2 \kms. The $T_{\rm mb}$ sensitivity is 0.04\,K. To probe similar scales among different isotopologues, the observations were convolved to an angular resolution of 31${}^{\prime\prime}$ (FWHM) from a resolution of 27\farcs1 for $^{12}$CO(2--1) and 28\farcs4 for $^{13}$CO (2--1) and C$^{18}$O (2--1), respectively. 

For G21.347$-$0.629 and G29.957$-$0.018, we used archival CO observations from three surveys. ${\rm {}^{13}CO}$(1--0) observations are obtained from the GRS survey \citep{JacksonRathborne:2006aa}. The GRS survey maps have a half-power beam width of 46\arcsec, and we correct all data for a main-beam efficiency of $\eta_{\rm mb} = 0.48$. Also, we used maps of ${\rm {}^{12}CO}$(1--0) and ${\rm C{}^{18}O}$(1--0) emission from the FOREST Unbiased Galactic plane Imaging survey with the Nobeyama 45-m telescope (FUGIN) at 0.65\,\kms\ spectral resolution (\citealt{UmemotoMinamidani:2017aa}), after convolving them from their original angular resolution of 20${}^{\prime\prime}$ (FWHM) to 46\arcsec.

\begin{table}
\centering
\caption{Spectral lines discussed in this work.}
\label{tbl:frequencies}
\begin{tabular}{lr}
\hline
\hline
Line  & Frequency [GHz]\\
\hline
OH ${}^2$$\Pi_{3/2}$ ($J$=3--2, $F$=1${}^{+}\! -\!2{}^{-}$) & 1.612231  \\
OH ${}^2$$\Pi_{3/2}$ ($J$=3--2, $F$=1${}^{+}\! -\!1{}^{-}$) &  1.665402  \\
OH ${}^2$$\Pi_{3/2}$ ($J$=3--2, $F$=2${}^{+}\! -\!2{}^{-}$) & 1.667359  \\
OH ${}^2$$\Pi_{3/2}$ ($J$=3--2, $F$=2${}^{+}\! -\!1{}^{-}$) & 1.720530 \\
\ion{H}{i} (${}^2$S$_{1/2}$, $F$=$1\! -\!0$) & 1.420406 \\ 
${}^{12}$CO ($J$=$1\! -\!0$)   & 115.2712  \\
${}^{12}$CO ($J$=$2\! -\!1$)   & 230.5380  \\
${}^{13}$CO ($J$=$1\! -\!0$   & 110.2013  \\
${}^{13}$CO ($J$=$2\! -\!1$)   & 220.3987  \\
C${}^{18}$O ($J$=$1\! -\!0$   & 109.7822 \\
C${}^{18}$O ($J$=$2\! -\!1$)   & 219.5604  \\
{[\ion{C}{ii}]} (${}^2$P$_{3/2}\! -\!{}^2$P$_{1/2}$) & 1900.537  \\
{[\ion{O}{i}] 63\,$\mu$m} (${}^3$P$_{1}\! -\!{}^3$P$_{2}$) & 4744.777  \\
{[${}^{13}$\ion{C}{ii}]} (${}^2$P${}_{3/2}\! -\!{}^2$P$_{1/2}$, $F$=$1\! -\!1$) &  1900.136 \\
{[${}^{13}$\ion{C}{ii}]} (${}^2$P$_{3/2}\! -\!{}^2$P$_{1/2}$, $F$=$2\! -\!1$) &  1900.466 \\
{[${}^{13}$\ion{C}{ii}]} (${}^2$P$_{3/2}\! -\!{}^2$P$_{1/2}$, $F$=$1\! -\!0$) &  1900.950 \\
\hline
\end{tabular}
\end{table}

\section{Analysis} \label{sec:analysis}

\subsection{OH and \ion{H}{I} column densities} \label{sec:analysis:oh}
We determined the column density of OH from the 1667\,MHz line, as its optical depth tends to be higher than the 1665\,MHz line, thus likely being detected at a better signal-to-noise for a given excitation temperature. We estimated the optical depth of the \ion{H}{i} and OH lines using
$\frac{F_{\rm line}}{F_{\rm cont}} = e^{-\tau}$, 
with $F_{\rm line}$ corresponding to the flux density of the spectral line without continuum subtraction, $F_{\rm cont}$ the continuum flux density and $\tau$ the optical depth. The continuum was derived from the absorption-free channels. Given the strong continuum, we assumed that \ion{H}{i} and OH emission is a negligible contribution to the spectrum after filtering extended emission with the interferometer. For saturated \ion{H}{i} absorption channels, we calculate a lower limit on the optical depth by assuming a minimum \ion{H}{i} absorption depth at the 3-$\sigma$ noise level.

The OH excitation temperature is typically within a few Kelvin of the cosmic microwave background temperature \citep{LiTang:2018ab,HafnerDawson:2023cg}, {with a distribution up to higher values  found up to $T_{\rm ex} = 20\,{\rm K}$}. We fix the excitation temperature at $T_{\rm ex} = 5\,{\rm K}$ and account for an uncertainty of a factor of two for the excitation temperature in the error analysis. The OH column density is then calculated as: 

\begin{equation} \label{eq:coldenseoh}
\frac{N_{\rm OH}}{T_{\rm ex}} = \frac{C_0}{f}\int{\tau \, dv}, 
\end{equation}
where $N_{\rm OH}$ describes the total OH column density in particles ${\rm cm}^{-2}$ integrated over the optical depth profile in velocity, $T_{\rm ex}$ is the
excitation temperature in Kelvin, and the pre-factors are given as $C_0 = 4.0\times10^{14} \,{\rm cm}^{-2} {\rm
 {\,}K}^{-1}{\rm km^{-1}s}$ for the 1665{\,}MHz line and $C_0 =
2.24\times10^{14} \,{\rm cm}^{-2} {\rm{\,}K}^{-1}{\rm km^{-1}s}$ for the
1667{\,}MHz line \citep[e.g.,][calculated using Einstein coefficients from
\citealt{Turner:1966aa}]{Goss:1968aa,TurnerHeiles:1971aa,StanimirovicWeisberg:2003aa}. For the filling factor we assumed $f=1$.

Similarly, we derive the \ion{H}{i} column densities using
\begin{equation} \label{eq:coldensehi}
N_\ion{H}{i} = 1.8224\times10^{18}\frac{T_{\rm spin}}{\rm K} \int{\tau({\rm v})\left(\frac{\rm dv}{\rm km\,s^{-1}}\right)}\,{\rm cm^{-2}}.
\end{equation}
We assume an average spin temperature of the absorbing \ion{H}{i} gas of $T_{\rm spin} = 100$\,K, as the absorption is dominated by the cold neutral medium. We estimated the uncertainties in this assumption to be about a factor of two, given the significant spread in the distribution found in previous studies \citep[e.g.,][]{HeilesTroland:2003aa,MurrayStanimirovic:2018ut,DickeyDempsey:2022fn}. 

We use $N_{\rm OH}$ and $N_{\rm HI}$ for further absorption features from the THOR survey, and adopt a uniform excitation temperature for $N_{\rm OH}$ of $T_{\rm ex} = 5$\,K, for uniformity with the data presented here.

\begin{figure*}
\centering
  \includegraphics[width=0.93\textwidth, trim=0 0 0 0]{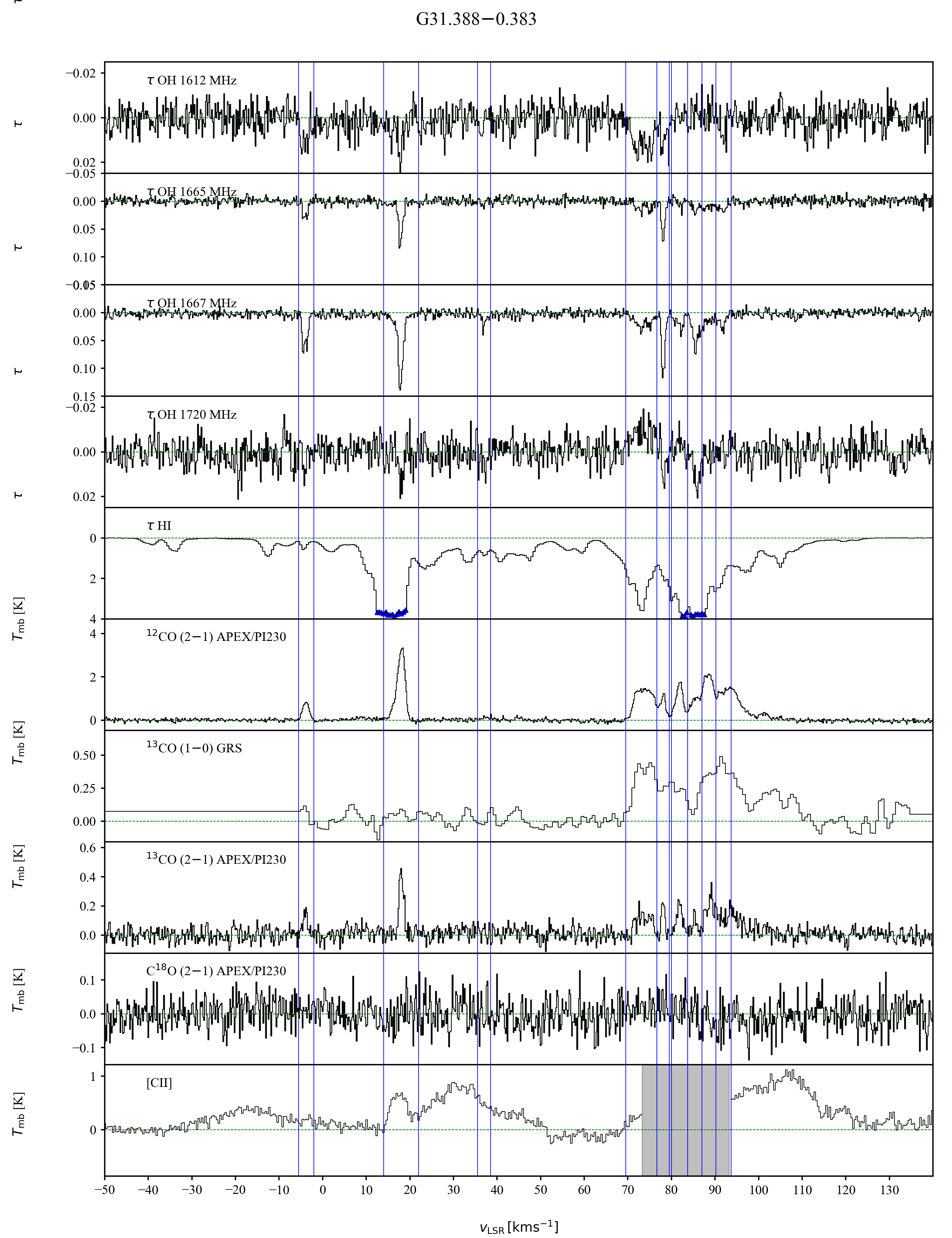}
\caption{
OH and \ion{H}{i} absorption towards the extragalactic source G31.388$-$0.383. The first five panels show optical depths of the four OH lines and \ion{H}{i}, as derived from the line-to-continuum ratio. For the \ion{H}{i} absorption, blue triangles indicate saturated channels. Emission spectra of the three CO isotopologues are shown below that. The data is taken with APEX and from the GRS \citep{JacksonRathborne:2006aa} survey. The bottom most panel displays the SOFIA/upGREAT observations of the [\ion{C}{ii}] 158\,$\mu$m line. Integration limits are shown as blue vertical lines. To emphasize that the OH and the \ion{H}{i} lines are observed towards strong background sources, and are - with the exception of stimulated emission - seen in absorption, the axis of the optical depth is inverted, i.e., goes from positive to negative values.}
  \label{fig:overview_transitions_g31}
\end{figure*}

\subsection{Molecular hydrogen column densities} \label{sec:analysis:co}
For all but one line-of-sight OH absorption feature we detect $^{12}$CO rotational lines (see Table~\ref{tbl:oh_integrals}), and use these to infer the molecular gas content. We convert integrated $^{12}$CO emission to molecular gas column density with a standard conversion factor of $\alpha_{\rm CO} = 2\times10^{20}$\,cm$^{-2}$ \citep{BolattoWolfire:2013aa}. We rescale the $^{12}$CO $J=2-1$ observations by a typical line ratio of $R_{21/10} = 0.7$ \citep[][]{SakamotoHasegawa:1997qk,PenalozaClark:2018im,den-BrokChatzigiannakis:2021mg}. 

While a large fraction of the line-of sight OH absorption features show detections in the $^{13}$CO (1--0) line, we decide to rely exclusively on a $^{12}$CO conversion factor to derive molecular gas column densities. This choice was made as the rotational levels of weak $^{13}$CO (1--0) detections are likely sub-thermally excited \citep[e.g.,][]{HeyerWilliams:2006cx}. Previous studies have found that the molecular gas column densities derived from $^{13}$CO emission at low densities may be underestimated when assuming local thermal equilibrium (LTE) conditions \citep{GoldsmithHeyer:2008aa,HeyerKrawczyk:2009yl,HeidermanEvans:2010lf}. We tested deriving $N_{\rm H_2}$ from the GRS $^{13}$CO (1--0) line emission reported here under LTE conditions (assuming $T_{\rm ex} = 10$\,K), and found that $N_{\rm H_2}$ was systematically lower by a factor of 2 or more for most but the highest column densities when comparing it to $N_{\rm H_2}$ derived directly from $^{12}$CO emission. We hence adopt $N_{\rm H_2}$ derived from $^{12}$CO emission. The integrated emission from all spectral lines is reported in Table~\ref{tbl:oh_integrals} to enable more refined modeling in future studies {\citep[e.g.,][]{RoueffGerin:2021qm}}.

For comparison with the THOR survey, we retrieved ${\rm {}^{12}CO}$(1--0) spectra from the FUGIN survey to update the $N_{\rm H_2}$ estimates for the following line-of-sight absorption features from \citet{RugelBeuther:2018aa}: THOR+28.806+0.174+79.5, THOR+31.242-0.110+79.5, THOR+31.242-0.110+83.5, THOR+39.565-0.040+24.0, and THOR+42.027-0.604+64.5. As described in  \citet{RugelBeuther:2018aa}, all data were convolved to 46$^{\prime\prime}$ before extracting the spectra at the source position. 

The $N_{\rm H_2}$ estimates for OH absorption features associated with \ion{H}{ii} regions were adopted from \citet{RugelBeuther:2018aa}, which used an excitation temperature of $T_{\rm ex} = 20$\,K for the GRS $^{13}$CO (1--0) transition, applied an optical depth correction, and a standard conversion factor of $N$(H$_2$) $\sim 3.8 \times 10^5$ $N$($^{13}$CO) \citep[][and references therein]{PinedaCaselli:2008aa,Dickman:1978lh,BolattoWolfire:2013aa}. Given that the conversion does not account for the varying isotope ratio of $^{12}$C/$^{13}$C with Galactocentric radius, we used the relation presented in \citet{JacobMenten:2020fo} to rescale $N_{\rm H_2}$ for OH absorption features within the solar circle. The Galactocentric radius was derived using parameters of the Galactic rotation curve from \citet{ReidMenten:2019aa}, as used in the bayesian distance calculator\footnote{v2.4.1} \citep{ReidDame:2016aa}. As the feature at 99.0\,\kms\ for G29.935$-$0.053 presented in this work is associated with an \ion{H}{ii} region, we used the GRS $^{13}$CO (1--0) emission, following the same procedure as described above. We exclude any OH absorption associated with the \ion{H}{ii} in G29.957$-$0.018, as the OH 1665 MHz transition is showing maser emission, which may affect the overall level population of the OH hyperfine splitting levels. After careful examination of the data, we used the ${\rm {}^{12}CO}$(1--0) emission as described above also for THOR+32.272-0.226+22.5, as the column density from GRS $^{13}$CO (1--0) was severely underestimated. 

\citet{BolattoWolfire:2013aa} estimate the uncertainties for $\alpha_{\rm CO}$ as 30\%. Given that $\alpha_{\rm CO}$ and $R_{21/10}$ may vary across the Milky Way, in particular in the diffuse ISM, the actual uncertainties in our sample are likely higher. We adopt an uncertainty for $N_{\rm H_2}$ of a factor of 2.

\section{Results} \label{sec:results}

\begin{figure*}
\centering
  \includegraphics[width=1\textwidth]{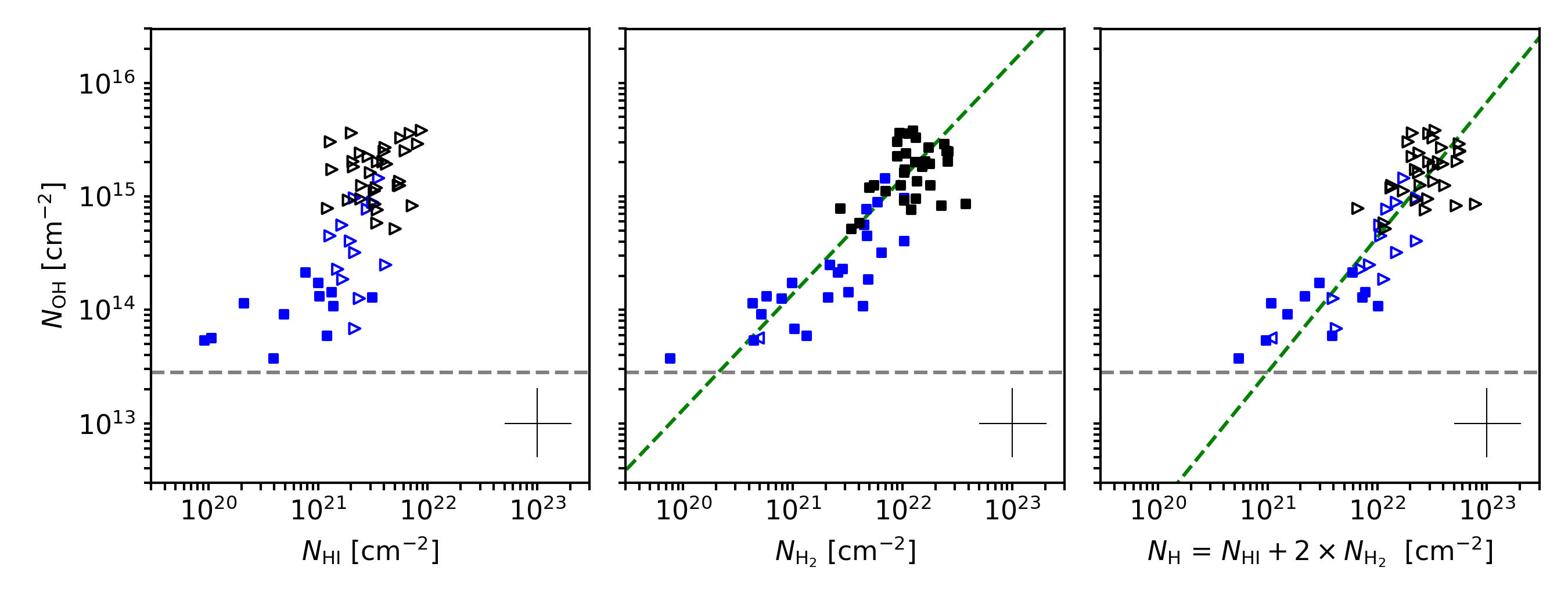}
\caption{OH column density vs. the column density of \ion{H}{i} ({\it left}), H$_2$ ({\it middle}), and the column density of the total number of hydrogen nuclei ({\it right}). Features associated with line-of-sight absorption are shown in {\it blue}, features associated with \ion{H}{ii} regions in black \citep{RugelBeuther:2018aa}. The middle panel shows $N_{\rm OH}$ vs. $N_{H_2}$. Lower limits in case of saturated \ion{H}{i} absorption in the {\it left} and {\it right} plots are indicated by triangles pointing to the right. There is one upper limit due to the non-detection of CO, which is indicated by triangles pointing to the left. Typical error bars are indicated in the lower right corner. The limit of a 3\,$\sigma$ detection is indicated by a dashed gray line. The green dashed line indicates the fit of a correlation to the data.}
  \label{fig:noh_vs_nh}
\end{figure*}

\begin{figure*}
\centering
  \includegraphics[width=1\textwidth]{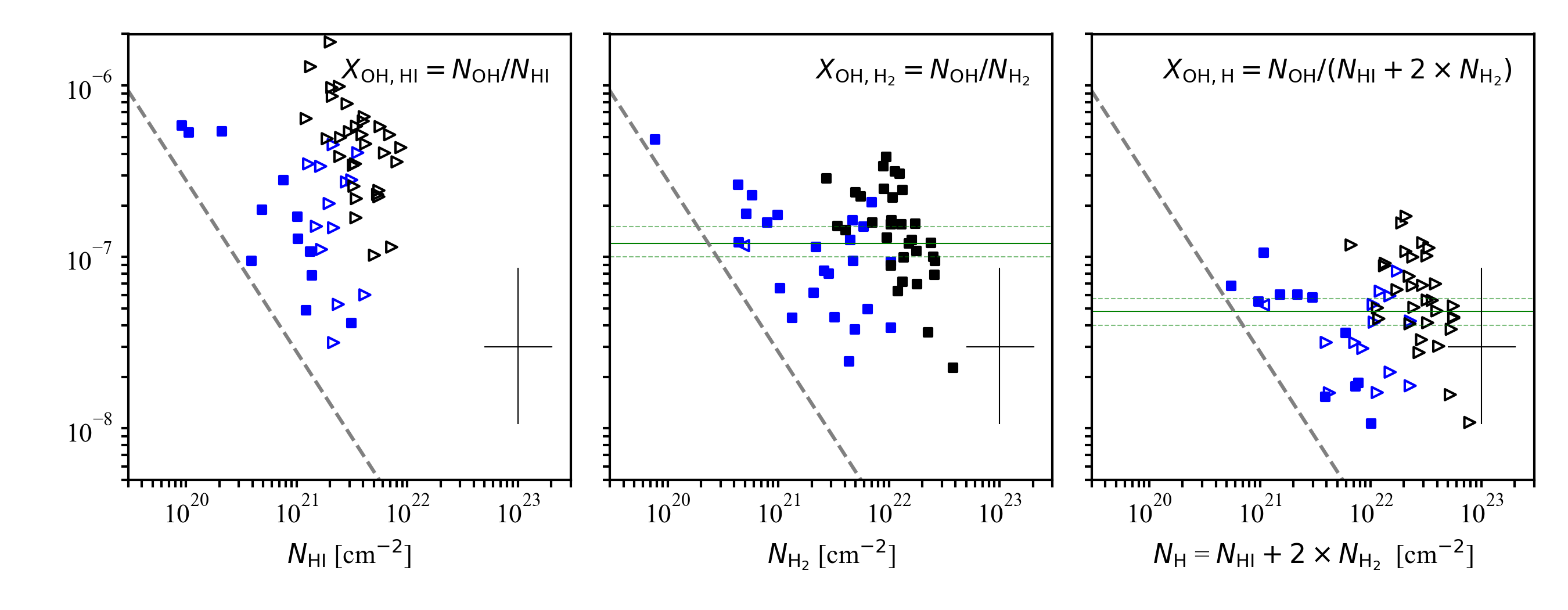}
\caption{Same as Fig.~\ref{fig:noh_vs_nh}, but for the column density-averaged OH abundance. The dashed gray line indicates 3\,$\sigma$ sensitivity limits of the OH absorption observations presented here. The green lines indicate the median abundance ({\it solid}) and 1$\sigma$ uncertainties ({\it dashed}).}
  \label{fig:xoh_vs_nh}
\end{figure*}

\subsection{OH line-of-sight absorption} \label{sec:results:losabsorption}

All but one of the OH absorption features have been detected in $^{12}$CO, 18/23 in $^{13}$CO, and 5/23 in C$^{18}$O. Detection of $^{12}$CO alone typically indicates gas at lower densities, while the detection of C$^{18}$O requires dense gas. 3/23 features are associated with molecular gas in the vicinity of \ion{H}{ii} regions (see Table~\ref{tbl:oh_integrals}), the other 20/23 features represent line-of-sight absorption. To derive column densities we measured integral properties of all spectral lines over 23 different intervals in velocity. 
An example for OH in low-density environments is the component at $\varv=37.0$\,\kms\ towards the extra-galactic source G31.388$-$0.383. It is only weakly detected in $^{12}$CO (2--1) emission. An example for OH absorption associated with dense gas as traced by C$^{18}$O is the deep OH absorption component at 56.5\,\kms\ towards G21.347$-$0.629. The \ion{H}{i} profile is also saturated here.
We detect super-position of velocity components of low- and high-density gas: The OH 1667 MHz absorption feature along the line of sight towards the extra-galactic source G21.347$-$0.629 at 55\,\kms\ (Fig.~\ref{fig:overview_transitions_g21}) shows a combination of a narrow component ($\Delta\varv\sim$0.9\,\kms) with a broader component ($\Delta\varv\sim$6\,\kms), with their velocity centroids clearly shifted with respect to each other. The narrow and deeper component is detected in all CO isotopologues, and its detection in C$^{18}$O clearly indicates the presence of dense gas in this narrow component. The broad component is detected only in the $^{12}$CO and $^{13}$CO lines, which indicates gas of lower density. Towards $\varv=$7\,\kms, a narrow component with a linewidth of 0.7\,\kms\ in OH 1667 MHz absorption and $^{13}$CO emission is set on top of a broader component with a line width of 5.7\,\kms, that is weak but detected in OH 1667 MHz absorption and readily visible in $^{12}$CO emission.

Population inversion is a common feature in ground state HFS transitions of OH, particularly, in its satellite lines at 1612 and 1720 MHz. We detected 6/23 features simultaneously in both satellite lines, of which three show conjugate behavior of stimulated emission in the OH 1612 MHz line and absorption in the OH 1720 MHz line. One shows the reverse configuration of absorption in the OH 1612 MHz line and stimulated emission in the OH 1720 MHz line, and two show absorption in both main lines. The remaining features show typically weak absorption in either the 1612 MHz or the 1720 MHz line at the detection limit. Our sample does not contain clearly `flipped' features which transition from emission to absorption, or vice versa, with increasing velocity, as has been described by \citet{HafnerDawson:2020qw}. We note conjugated satellite line peaks towards G29.935$-$0.053 at $\varv=$~101.2\,\kms, which are shifted from the peak velocity of the main lines at 99\,\kms. However, we do not significantly detect a reversed feature at lower velocities. Figure~\ref{fig:overview_transitions_g21} indicates a potential reversal of the satellite lines across the broad velocity component at $v=$~7\,\kms\ towards G21.347$-$0.629. 

An example of OH main line inversion, which is typically seen in high-mass star forming regions \citep[e.g.,][]{BeutherWalsh:2019ng} pumped by strong infrared emission \citep[e.g.,][]{GuibertRieu:1978aa} can be seen towards G29.957$-$0.018, in the components around 100\,\kms\ which is associated with an UC\ion{H}{ii} region. Maser emission in the OH 1665 MHz line is visible between 100--105\,\kms. Both main lines show maser emission at $\sim$107\,\kms (Fig. \ref{fig:overview_transitions_g29}). 

To summarize, we find that absorption in the OH 1665 MHz and OH 1667 MHz lines occurs in diffuse to dense environments. In the following sections, we discuss the OH abundance in these different ISM conditions.

\subsection{The OH abundance in atomic and molecular gas} \label{sec:results:abundance}
Given the occurrence of OH absorption in different ISM conditions noted in Sect.~\ref{sec:results:losabsorption}, this section analyzes the relation between OH absorption and tracers of different ISM gas phases. We trace the atomic gas component with the simultaneously recorded \ion{H}{I} 21\,cm line absorption. To trace phases dominated by molecular hydrogen we make use of  new and archival observations of CO. Both approaches have limitations - the \ion{H}{I} 21\,cm line saturates at high column densities, while CO may be less abundant or not present at all at low molecular hydrogen abundances. As CO emission is one of the most commonly used tracer of molecular gas in the literature, we adopt CO as our reference tracers of molecular hydrogen, and discuss limitations and implications below.

The probed ranges of $N_{\rm OH}$ and $N_{\rm H_2}$ extend to lower column densities than in the THOR sample. The OH observations in this study extend to column densities as low as 3.7$\times10^{13}$\,cm$^{-2}$ and up to 8.5$\times10^{14}$\,cm$^{-2}$. After rescaling $N_{\rm OH}$ to the same excitation temperature as the data presented here, the OH observations in THOR\footnote{OH absorption detections presented here were not removed from the original sample presented in \citet{RugelBeuther:2018aa}}\citep{RugelBeuther:2018aa} probed higher column densities between 1.2$\times10^{14}$ and 3.8$\times10^{15}$\,cm$^{-2}$. The column densities of molecular hydrogen as probed in this work by CO emission extend from $N_{\rm H_2}$=7.7$\times10^{19}$\,cm$^{-2}$ to 6.0$\times10^{22}$\,cm$^{-2}$, whereas the OH detections reported in \citet{RugelBeuther:2018aa} occur at $N_{\rm H_2}$=7.9$\times10^{20}$\,cm$^{-2}$ to 4.7$\times10^{22}$\,cm$^{-2}$.  

In the following, we will jointly analyze both datasets. Due to higher sensitivities and higher spectral resolution (channel width of 0.2\,\kms\ vs. 1.5\,\kms\ for THOR) we are able to identify more narrow features in the new sample presented here. 

Figure~\ref{fig:noh_vs_nh} compares the OH column densities to the column density of atomic hydrogen and CO-derived molecular hydrogen, as well as to the total column density of hydrogen nuclei, which is given as $N_{\rm H}$ = 2$\times$$N_{\rm H_2}$ + $N_{\rm HI}$. Comparing with the column density of atomic hydrogen, we see that the \ion{H}{i} absorption is saturated for all OH features with $N_{\rm HI}>1.4\times10^{21}$cm$^{-2}$. We identify 12 features in the dataset presented in this work for which the \ion{H}{i} absorption is not saturated. 

The middle panel compares $N_{\rm OH}$ with the column density of molecular hydrogen as traced by CO. The ${\rm H_2}$ column densities are estimated from different isotopologues as well as different rotational transitions of CO (see Sect.~\ref{sec:analysis:co}). One feature in the new dataset does not have a counterpart in any CO {line}, namely the feature at 68.8\,\kms\ toward G21.347$-$0.629. To investigate the relationship between OH and molecular hydrogen column densities, we fit the data with a power-law. To improve on previous estimates in \citet{RugelBeuther:2018aa} and to mitigate biases towards sub-linear trends, we fit a slope in logarithmic space with a least squares algorithm while properly accounting for uniform errors in both x and y directions, respectively \citep{York:1966cc}. We find the quantities to be related by a power law $N_{\rm OH} \propto N_{\rm H_2}^{m_{\rm H_2}}$ with index $m_{\rm H_2}=1.0\pm{0.1}$. In the panel on the right-hand side of Fig.~\ref{fig:noh_vs_nh}, we compare the OH column density to the total column density of hydrogen nuclei. Since the \ion{H}{i} column density saturates in many instances, also $N_{\rm H}$ is a lower limit. If we took only the 12 data points for which $N_{\rm \ion{H}{i}}$ is not saturated, the Spearman's rank coefficient is $r_s=0.6$ with $p=0.08$, which indicates that we cannot reject the null hypothesis at the present sample size. If we assume the contribution of $N_{\rm H_2}$ to $N_{\rm H}$ to dominate in most cases where $N_{\rm \ion{H}{i}}$ saturates, we can also consider $N_{\rm H}$ estimates with saturated \ion{H}{i} absorption. In this case we find an exponent for a power-law relation between $N_{\rm OH}$ and $N_{\rm H}$ of $m_{\rm H}=1.2\pm0.1$. 

\begin{figure}
\centering
\includegraphics[width=1\columnwidth]{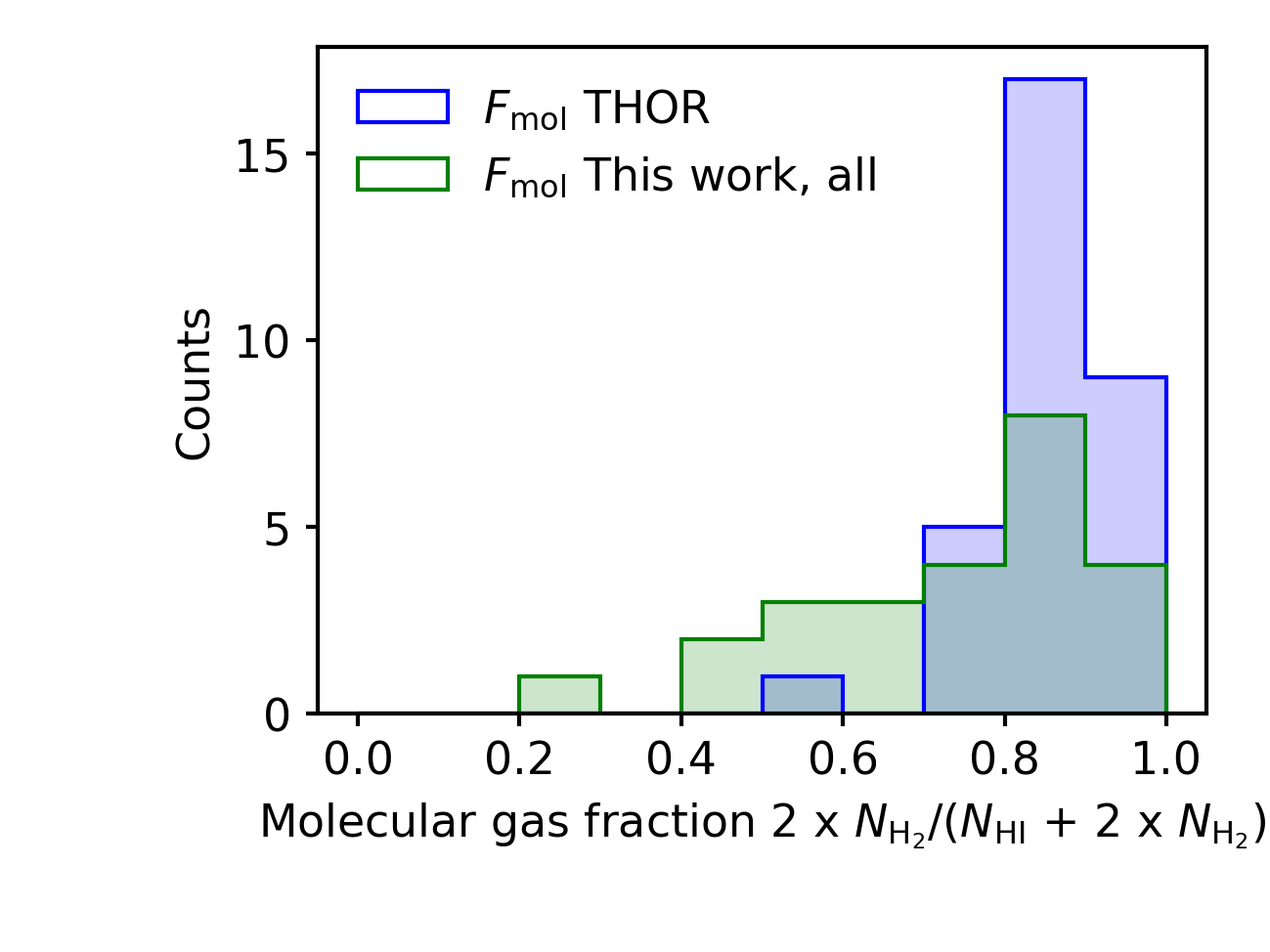}
\caption{Distribution of molecular gas fraction in this work ({\it green}) in comparison to the sample from THOR ({\it blue}). Many of the features included in both samples are upper limits on the molecular gas fraction, as the \ion{H}{i} absorption saturate. We hence highlight the sub-sample of features with non-saturating \ion{H}{i} absorption presented in this work ({\it red}).}
\label{fig:fmol_thorhr_thor}
\end{figure}

We investigate the OH abundance with respect to atomic, molecular and total hydrogen column density in Fig.~\ref{fig:xoh_vs_nh}. The median OH abundance with respect to molecular hydrogen of $X_{\rm OH, H_2} = N_{\rm OH}/N_{\rm H_2} = 1.2^{+0.3}_{-0.2}\times10^{-7}$. The upper and lower bounds indicate 16\% and 84\% confidence intervals of the abundance after bootstrapping the posterior distribution of the median over the error margins indicated in Sect.~\ref{sec:analysis:co} (0.3 dex in $N_{\rm OH}$, $N_{\rm H_2}$, and $N_{\rm \ion{H}{i}}$). The median abundance with respect to the total column density of hydrogen nuclei is $X_{\rm OH, H} = N_{\rm OH}/N_{\rm H}= 3.8^{+2.0}_{-1.4}\times10^{-8}$, omitting features with saturated \ion{H}{i} lines. The sample with non-saturated \ion{H}{i} column densities spreads a large range in molecular gas fractions (see Fig.~\ref{fig:fmol_thorhr_thor}). This is holds true if we assume a lower spin temperature for $N_{\rm \ion{H}{i}}$ (Fig.~\ref{fig:fmol_thorhr_thor_50k}). Sources with saturated \ion{H}{i} column densities tend to occur in line-of-sight features with high molecular gas fraction, and the lower limits of $N_{\rm \ion{H}{i}}$ would need to underestimate it by a factor of a few to significantly lower the molecular gas fractions. Including all estimates of $N_{\rm H}$, the median abundance is $X_{\rm OH, H} =4.8^{+0.9}_{-0.8}\times10^{-8}$ , which is only slightly higher than the result with \ion{H}{i} detections only. 

\begin{figure}[!ht]
\centering
  \includegraphics[width=1\columnwidth]{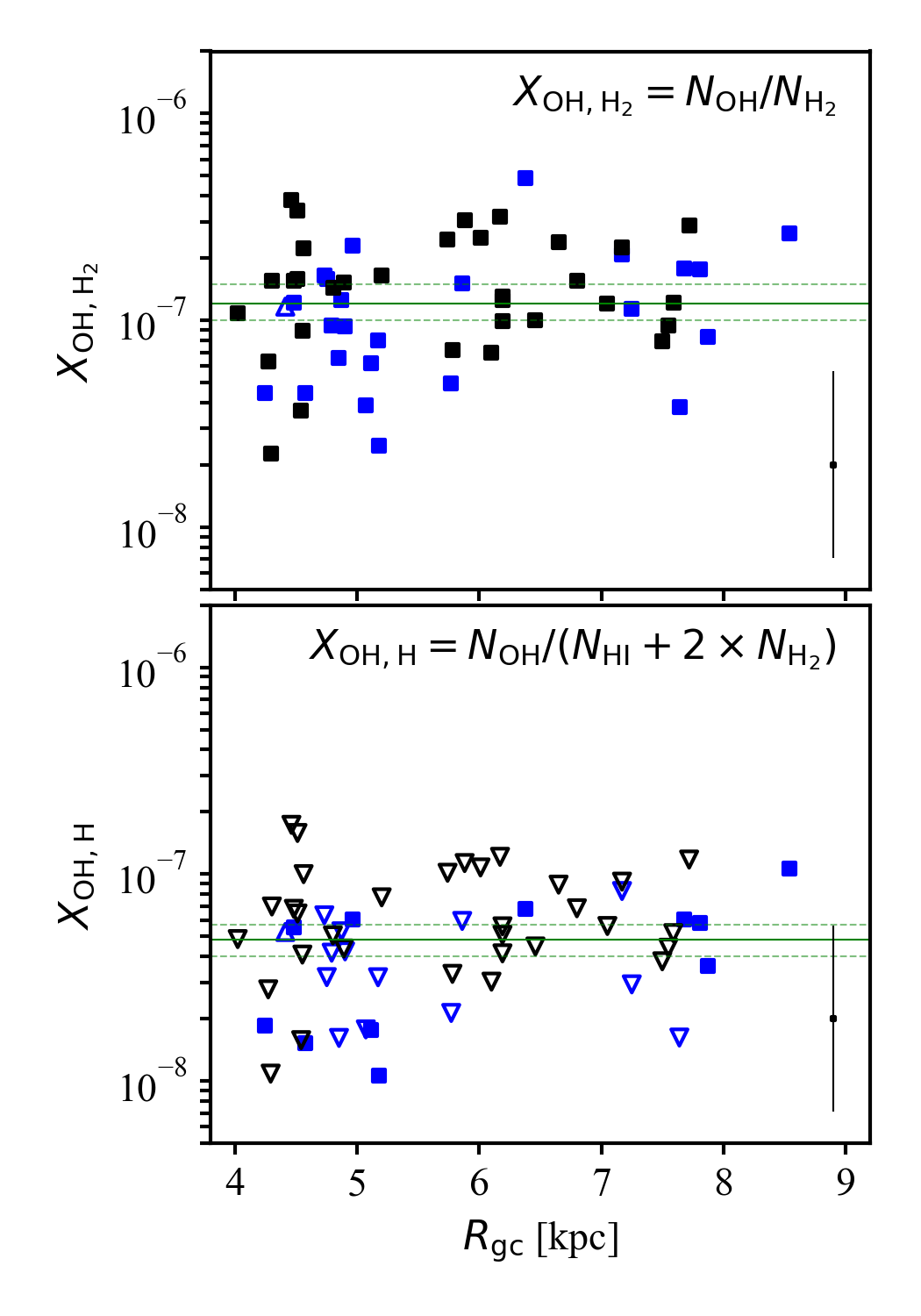}
\caption{Same as Fig.~\ref{fig:xoh_vs_nh} but for $X_{\rm OH, H_2}$ and $X_{\rm OH, H}$  vs. $R_{\rm GC}$.}
  \label{fig:xoh_vs_rgc}
\end{figure}

The abundances for line-of-sight components are $X_{\rm OH, H_2} =  1.1^{+0.7}_{-0.5}\times10^{-7}$ and $X_{\rm OH, H} = N_{\rm OH}/N_{\rm H}= 3.8^{+2.1}_{-1.4}\times10^{-8}$. This is slightly lower than the abundances for OH associated with \ion{H}{ii} regions are $X_{\rm OH, H_2} = 1.4^{+3.7}_{-2.9}\times10^{-7}$ and $X_{\rm OH, H} = 5.9^{+1.4}_{-1.2}\times10^{-8}$. Figure~\ref{fig:xoh_vs_rgc} shows the molecular and total hydrogen abundance vs. Galactocentric radius. While slight variations may be present, the Spearman's rank test yielded $p>0.1$ for any correlation between {$X_{\rm OH, H_2}$ or $X_{\rm OH, H}$} and $R_{\rm GC}$, both for line-of-sight components and molecular clouds associated with \ion{H}{ii} regions individually, as well as all data combined.

\subsection{Detections of the [\ion{C}{ii}] 158\,$\mu$m emission towards OH absorption components} \label{sec:results:cii_g31}
The [\ion{C}{ii}] observations for G31.388$-$0.383 and G29.935$-$0.053 are shown Figs.~\ref{fig:overview_transitions_g31}~and~\ref{fig:overview_transitions_g29p935}, respectively. We find widespread [\ion{C}{ii}] emission in both spectra originating from different velocity components along the line of sight. Given its strength and velocity, the main component of the [\ion{C}{ii}] emission in G29.935$-$0.053 is dominated by PDR emission from the \ion{H}{ii} region complex. G31.388$-$0.383 shows enhanced [\ion{C}{ii}] emission around 100\,\kms\, and is affected by contamination in the reference position between 73 and 93\,\kms (Fig.~\ref{fig:g31p388_cii}). Below, we discuss [\ion{C}{ii}] emission as it occurs in conjunction with OH main line absorption. 

For G29.935$-$0.053, Fig.~\ref{fig:overview_transitions_g29p935} does not show [\ion{C}{ii}] 158\,$\mu$m emission that can be uniquely associated with line-of-sight OH absorption. While the [\ion{C}{ii}] 158\,$\mu$m emission at $v=51$\,\kms\ is close in velocity to an OH absorption feature, its peak does not match, but rather seems to be associated with the deep \ion{H}{i} absorption feature, as it spans a similar range in velocity. The lower emission between 0 and 30\,\kms\ appears to be consistent between different observing dates and reference positions, and is seen in all of the pixels (Fig.~\ref{fig:g29p935_cii_zoomlow}). It could indicate absorption, although we note baseline ripples of similar strength. As a confident estimate of the continuum level was not possible, we do not discuss the feature here further. 

For G31.388$-$0.383 (Fig.~\ref{fig:overview_transitions_g31}), we see broad [\ion{C}{ii}] 158\,$\mu$m emission towards $v=37$\,\kms. However, we cannot identify a specific [\ion{C}{ii}] 158\,$\mu$m emission feature associated in velocity with the OH absorption. Towards $v=18$\,\kms\ we detect significant [\ion{C}{ii}] 158\,$\mu$m emission which is potentially associated with OH absorption. We discuss its gas content in the below. After removing contributions from the underlying broader emission profile, the component at 18\,\kms yields an integrated [\ion{C}{ii}] 158\,$\mu$m emission of 1.8 K\,\kms.

We estimate the column density of the [\ion{C}{ii}] 158\,$\mu$m emission, re-arranging equation 26 for the optically thin case from \citet{GoldsmithLanger:2012aa} (see also \citealt{BeutherSchneider:2022pt}): 
\begin{equation}
\begin{aligned}
N_{\rm CII} = & 2.9\times10^{15} \times \left[1+0.5{e}^{(91.25/T_{\rm kin})}\times\left(1+\frac{2.4\times10^{-6}}{n R_{ul}}\right)\right] \\
                      & \times \int  T_{\rm mb} d{v}\\
\end{aligned}
\label{eq:ncii}
\end{equation}
We assume a kinetic temperature of $T_{\rm kin} =$100\,K, a collision rate coefficient\footnote{\url{https://home.strw.leidenuniv.nl/~moldata/C+.html}} of $R_{\rm ul} = 5\times10^{-10}$ for collisions with ortho- and para-H$_2$ ($4.7\times10^{-10}$ and, respectively, $5.5\times10^{-10}$; \citealt{Schoiervan-der-Tak:2005aa,LiqueWerfelli:2013fx}) in an equal ratio, and densities of $n=10^3 $cm$^{-3}$. We find the column density to be $N_{\rm C^{+}} =4.3\times10^{16}\,{\rm cm}^{-2}$. 

In Sect.~\ref{sec:analysis:co}, we derived the H$_2$ column density from the integrated $^{12}$CO emission directly. Here, we estimate the $^{12}$CO column density from its $J$=2--1 emission with an excitation temperature of $T_{\rm ex} =$10\,K using equation 79 in \citet{MangumShirley:2015aa}, and obtain $N_{\rm ^{12}CO} =7.1\times10^{15}{\rm cm}^{-2}$. While we do not have tracers of neutral carbon available, this indicates considerable ionization fractions, consistent with models of translucent clouds \citep{van-DishoeckBlack:1988pc}. We note that 1667 MHz OH absorption appears to consist of multiple components. A narrow one, which is detected in both $^{12}$CO and $^{13}$CO emission, as well as a broader one, which is only detected in $^{12}$CO. The $^{13}$CO emission has a line width of 1.2\,\kms, which is considerably more narrow than the [\ion{C}{ii}] 158\,$\mu$m emission with a FWHM of 3.7\,\kms. It may be that OH is picked up from both a diffuse, broad component with high carbon ionization fraction, as well as denser, predominantly molecular component.  

\subsection{C$^+$ in the \ion{H}{ii} region complex G29.935$-$0.053}  \label{sec:results:cii_g29}

\begin{figure}
\centering
  \includegraphics[height=0.85\textheight]{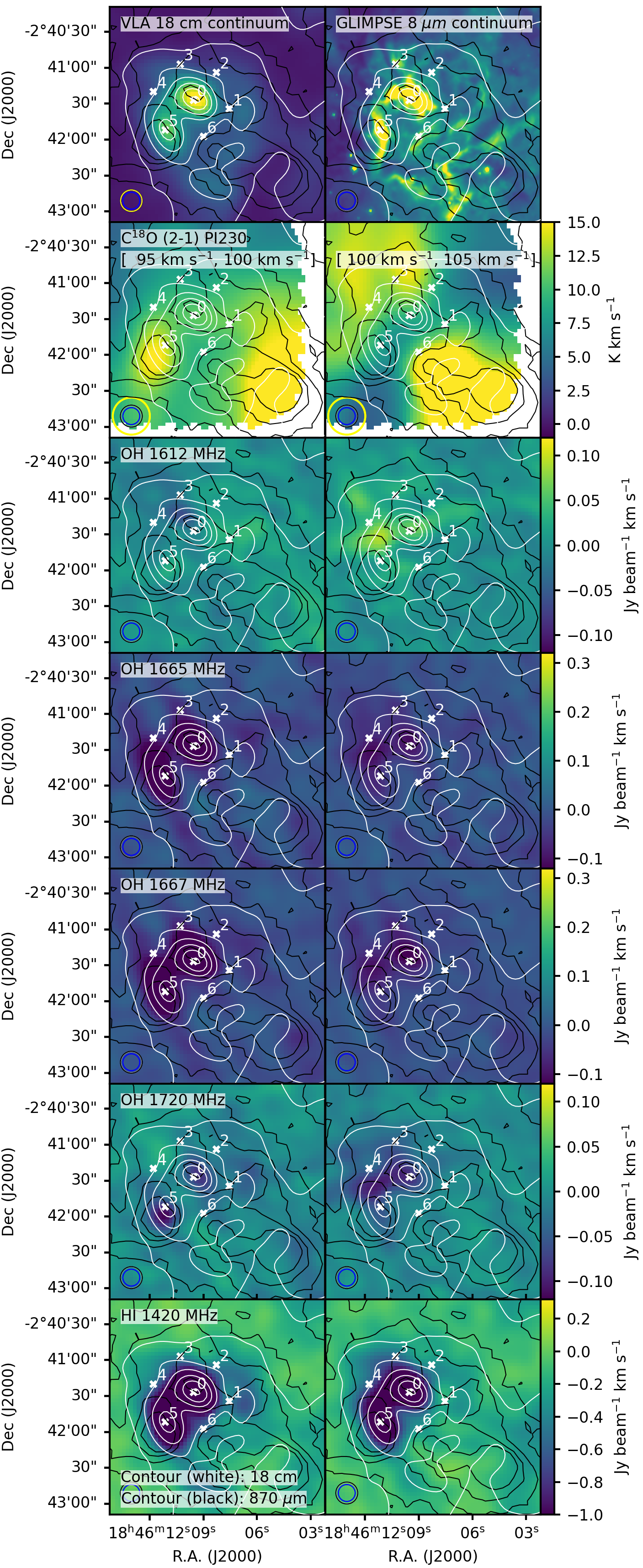}
\caption{Distribution of VLA 18\,cm and GLIMPSE 8 $\mu$m continuum emission \citep[top row][]{BenjaminChurchwell:2003aa} in G29.935$-$0.053, with contours of ATLASGAL 870\,$\mu$m emission \citep[black; at 0.5, 1.0, 1.5 and 2 \jyb][]{SchullerMenten:2009aa} and VLA 18\,cm continuum (white; in steps of 0.1\,\jyb, starting at 0.03\,\jyb). The beam of both observations is indicated with a black circle in the lower left. The positions of the upGREAT LFA pixels are indicated, and the beamsize indicated in blue. The following rows show spectral lines integrated over 95--100\,\kms\ and 100--105\,\kms\ in the left and right panels, respectively.}
  \label{fig:mommap_g29_all}
\end{figure}

The [\ion{C}{ii}] 158\,$\mu$m emission associated with G29.935$-$0.053 shows complex line shapes, potentially affected by self-absorption (Fig.~\ref{fig:overview_transitions_g29p935}). Figure~\ref{fig:mommap_g29_all} shows the radio continuum emission at 1.4 GHz in contours across G29.935$-$0.053, together with the positions observed during the SOFIA observations. Pixel 0 (see Sect.~\ref{sec:analysis:oh}) and pixel 5 are associated with peaks of the continuum emission, while the other pixels sample different parts of the HII region complex. The complete data are presented in Fig.~\ref{fig:g29_spec}. 

The [\ion{C}{ii}] 158\,$\mu$m emission at pixel 0 extends between 76\,\kms\ and 127\,\kms\ and contains three distinguishable emission peaks. The [\ion{O}{i}] 63\,$\mu$m emission shows a similar line shape (Fig.~\ref{fig:g29p935_oi}). Recent studies have interpreted [\ion{C}{ii}] 158\,$\mu$m and [\ion{O}{i}] 63\,$\mu$m emission from \ion{H}{ii} region complexes as a complex interplay of warm background emission and cold foreground absorption \citep{GuevaraStutzki:2020du,GuevaraStutzki:2024ss}, motivated by simultaneous detections of [$^{13}$\ion{C}{ii}] and the [\ion{O}{i}] 145 $\mu$m emission, which are typically optically thin and dominated by the warm background emission. No measurements of the [\ion{O}{i}] 145 $\mu$m emission are available for this study. A search for [$^{13}$\ion{C}{ii}] emission shows faint features at the correct velocities, but remains in the end inconclusive: The upGREAT SOFIA observations in the rest frame of the [$^{13}$\ion{C}{ii}] $F$=1--0 and [$^{13}$\ion{C}{ii}] $F$=1--1 lines (Fig.~\ref{fig:g29p935_cii_13cii}) indicates an emission peak at rest velocities of G29.935$-$0.053 in the [$^{13}$\ion{C}{ii}] $F$=1--0 line, but it is over-posed by potential absorption, baseline ripples, and Galactic emission. The [$^{13}$\ion{C}{ii}] $F$=1--1 line, located at lower frequencies than the [\ion{C}{ii}] 158\,$\mu$m line, is the weakest of the three hyperfine splitting transitions and only shows marginal emission.  Without a clear identification of the [$^{13}$\ion{C}{ii}] hyperfine-splitting transitions and no availability of the [\ion{O}{i}] 145 $\mu$m, the background and foreground components contributing to the [\ion{C}{ii}] 158\,$\mu$m and the [\ion{O}{i}] 63\,$\mu$m cannot be disentangled. 

We point out that the $^{12}$CO emission bears resemblance to the three-peak structure of the [\ion{C}{ii}] 158\,$\mu$m emission. To some extent this also holds for the $^{13}$CO emission. Both show a trough at similar velocities as the [\ion{C}{ii}] 158\,$\mu$m profile. The strongest peak of the C$^{18}$O emission is aligned with the trough in the $^{12}$CO and $^{13}$CO emission, indicating the presence of dense gas in this component, and suggesting that $^{12}$CO and $^{13}$CO lines are both optically thick at these velocities. This dense gas component also matches the velocity of the main line OH absorption peak. The C$^{18}$O line shows several other peaks, both at lower velocities towards 93\,\kms, and at higher velocities at 102\,\kms\ and 110\,\kms. The feature at 102\,\kms\ falls into the red tail of the OH main line absorption feature. Given the alignment of the OH absorption, the C$^{18}$O emission, and the trough in [\ion{C}{ii}] 158\,$\mu$m emission over many of the pixels of the SOFIA observation (Fig.~\ref{fig:g29_spec}), the trough in the [\ion{C}{ii}] 158\,$\mu$m and the [\ion{O}{i}] 63\,$\mu$m emission may corresponds to absorption of a colder foreground component. Additional observational constraints, however, are necessary to resolve the geometry of the region giving rise to the observed line profiles.

\subsection{Spatial and spectral variation of the OH excitation}
Resolved 18\,cm continuum observations in G29.935$-$0.053 enable us to investigate spatial and spectral variation of the excitation conditions of OH.  Figure~\ref{fig:mommap_g29_all} indicates clumps of dense gas as traced by 870\,$\mu$m emission from cold dust spatially surrounding 18\,cm continuum emission. The Galactic Legacy Infrared Midplane Survey Extraordinaire (GLIMPSE) 8\,$\mu$m map traces emission from poly aromatic hydrocarbons (PAHs) which appears to lie at interfaces between cm-continuum emission and dust emission, indicating the location of PDR regions. The continuum emission towards pixels 0, 2, 3, 4 appears to lie behind a layer of molecular and atomic gas, as OH and \ion{H}{I} is detected in absorption (Figure~\ref{fig:g29_spec}). Radio recombination lines at 5\,cm appear to peak redshifted (towards Pixels 2, 3, 4 and 5) or at similar velocities (Pixel 0) as the OH absorption, indicating expansion away in the back of the molecular gas emission. Pixel 1 and 6 do not show OH absorption, indicating that the continuum emission may lie in front of the bulk of the molecular gas. The dense molecular gas appears to be distributed in different spatial and spectral components across the region. The integrated C$^{18}$O (2--1) emission between $95-100$\,\kms\ appears to peak towards the 870\,$\mu$m emission peak, indicating that this velocity component may trace cold, dense molecular clumps. Emission at velocities between $100-105$\,\kms\ appears to extend towards the southwest and northeast of the \ion{H}{ii} region. 

The OH satellite lines peak at redder velocities than the minimum of the OH main line absorption and maximum of the C$^{18}$O emission. This indicates the presence of distinct velocity components contributing to the OH lines, as also the CO emission in Fig.~\ref{fig:mommap_g29_all} appears to originate from different spatial locations inside G29.935$-$0.053 for different velocity ranges. Also for the OH main lines, blueshifted components appear to be strongest towards the peak of the continuum emission and close to the peak of the 870\,$\mu$m emission. The redshifted component appears to originate offset towards the northeast. The satellite line absorption (1720 MHz) and emission (1612 MHz) appear to follow the distribution of the redshifted component of the OH main line absorption. 

\section{Discussion} \label{sec:discussion}

Section~\ref{sec:results:losabsorption} shows that gas at different densities appears to contribute to OH absorption, as inferred from the detection statistics of CO isotopologs. This is in agreement with findings in the literature: For example, OH has been found in PDRs at densities of $n_{\rm H_2}\sim10^{6}$\,${\rm cm}^{-3}$ (e.g., DR21; \citealt{JonesField:1994aa,Koley:2023ro}), and at intermediate to low densities in molecular cloud envelops of $n<10^{3}$\,${\rm cm}^{-3}$ (TMC-1; \citealt{EbisawaSakai:2019aa}). 
\citet{BuschAllen:2019cr} found an upper limit for the volume density of OH in CO-dark gas of $n<10^{2}$\,${\rm cm}^{-3}$ through chemical modeling, and \citet{BuschEngelke:2021nk} inferred extremely low average densities of {as low as} {$n\sim7\times 10^{-3}$\,${\rm cm}^{-3}$} for a layer of CO-dark OH gas in the Outer Galaxy. The consistent appearance of OH throughout very different environments is likely related to its natural appearance in oxygen chemistry \citep[e.g.,][and references therein]{van-DishoeckHerbst:2013aa}, and the HFS transitions of the rotational ground state of OH are easily excited due to their small energy difference. 

The initial motivation for this study was to follow-up and search for OH features which are candidates for tracing CO-dark gas as a potential tracer for the transition regions between atomic and molecular gas. Our investigations only revealed one feature without a simultaneous detection of carbon monoxide, at least within the sensitivities and angular resolution of the data at hand. This detection rate appears to be in agreement with previous studies: \citet{LiTang:2018ab} found two sources significantly above our 3$\sigma$ limit of 3-4$\times10^{13}$\,${\rm cm}^{-2}$ without a counterpart CO detection (there are 3 more at our 3$\sigma$ limit). Conversely, they also find several OH features at similar column densities which show CO counterparts. Both the sample presented here and the THOR sample indicate OH absorption features at intermediate molecular gas fractions ($<$0.8; Fig.~\ref{fig:fmol_thorhr_thor}), i.e., for which the hydrogen gas content is not fully dominated by molecular hydrogen. These may fall into the diffuse molecular interstellar cloud type \citep{SnowMcCall:2006aa}. A particularly interesting feature in this context is the OH absorption feature at 37\,\kms\ towards G31.388$-$0.383, which shows a highly elevated $X_{\rm OH, H_2}$ but an average $X_{\rm OH, H}$ value at $R_{\rm GC} = 6.4$\,kpc in Fig.~\ref{fig:xoh_vs_rgc}, indicating a large atomic gas fraction. We also note that $\alpha_{\rm CO}$ may be significantly underestimated at $N_{\rm H_2}=0.8\times10^{20}$\,cm$^{-2}$ due to low CO abundances \citep{van-DishoeckBlack:1988pc,Liszt:2007pd}.

Our findings of a linear relation between $N_{\rm OH}$, $N_{\rm H_2}$ and $N_{\rm H}$ is in agreement with previous studies. \citet{XuLi:2016aa} find a linear relation between emission in the OH 1665 MHz line and dust extinction in the Taurus molecular cloud. \citet{LisztLucas:1996aa} find OH absorption to be tightly correlated with HCO$^+$. Our results show OH abundances which agree well with other measurements within the uncertainties and the scatter of our sample \citep[e.g.,][]{Goss:1968aa,Crutcher:1979aa,JacobNeufeld:2022aa,AlbertssonIndriolo:2014aa}. While the median abundance is slightly lower in more diffuse line-of-sight clouds than in dense molecular clouds associated within \ion{H}{ii} regions, they agree well within errors. We do not detect significant abundance variations with Galactocentric radius within the Solar circle. The measured abundances agree well with chemical models, which predict abundances between $10^{-8} - 10^{-7}$ for $A_V<1$, which can increase to $\sim10^{-6}$ for $A_V>1$ \citep[e.g.,][]{HollenbachKaufman:2009aa,HollenbachKaufman:2012aa}. 

Towards high densities in star forming regions and in the presence of strong infrared radiation fields, the excitation structure of the OH hyperfine splitting transitions \citep[e.g.,][]{Nguyen-Q-RieuWinnberg:1976aa} may vary from diffuse clouds, as indicated by the maser emission in G29.957$-$0.018 around 100\,\kms. A combined modeling of the hyperfine splitting transitions of OH at 18\,cm with absorption of the OH rotational transitions into the ground state at far-infrared wavelengths \citep[e.g.,][]{WiesemeyerGusten:2012aa,JacobNeufeld:2022aa} can further constrain the excitation structure. For OH absorption and emission features associated with high-mass star formation, previous studies revealed significant pumping by the far infrared radiation field \citep[e.g.,][]{GoicoecheaCernicharo:2002sr,CsengeriMenten:2012aa}.

The analysis of the OH abundance presented here relies on CO as tracer of the bulk molecular gas, as measurements of CO are readily available and provide the necessary spectral information. However, we acknowledge that this introduces biases in the presented analysis: At the low end, it likely underestimates the true H$_2$ column density given the presence of significant fractions of CO-dark gas in diffuse ISM regions \citep[e.g.,][]{GrenierCasandjian:2005aa,PinedaLanger:2013aa}. This would imply slightly steeper relations than determined in Sect.~\ref{sec:results:abundance}. The abundance itself, however, does not show significant trends towards higher abundances at low H$_2$ column densities (Fig.~\ref{fig:xoh_vs_nh}) after accounting for the scatter and the approximate sensitivity limit (gray dashed line in Fig.~\ref{fig:xoh_vs_nh}). Indeed, if CO was underrepresenting the H$_2$ column density, an expectation could be that the OH abundance increases with increasing Galactocentric radius as we derive $N_{\rm H_2}$ from the CO emission. While \citet{PinedaLanger:2013aa} indicate an increase in the CO-dark gas fraction by a factor of $\sim$3 between $R_{\rm GC} = 4$\,kpc and $R_{\rm GC} = 8$\,kpc, our data do not show evidence for a significant trend of the OH abundance with Galactocentric radius, neither for OH in line-of-sight absorption features, nor for OH associated with star formation (Fig.~\ref{fig:xoh_vs_rgc}). This indicates that potential contributions of CO-dark gas to $N_{\rm H_2}$ may be of limited impact, at least for the data presented here. Nevertheless, a comparison also to more suitable molecular gas tracers in the diffuse ISM, i.e., 3D dust maps \citep[e.g.,][]{MullensZucker:2024xs}, HCO$^+$ \citep[e.g.,][]{RybarczykStanimirovic:2022kt,KimSchilke:2023sa} or hydride absorption \citep[e.g.,][]{JacobNeufeld:2022aa}, may be beneficial to further assess the OH abundances in the regime where CO ceases to be a reliable molecular gas tracer. 

While the [\ion{C}{ii}] 158\,$\mu$m line has been used as a tracer of CO-dark gas \citep[e.g.,][]{PinedaLanger:2013aa}, not all of the emission is molecular-gas dominated \citep[e.g.,][]{FraneckWalch:2018ab,EbagezioSeifried:2024xt}. As we find OH in lines of sight which have low molecular gas fractions (Sect.~\ref{sec:analysis:oh}), we note at the same time for G29.935$-$0.053 and G31.388$-$0.383 that not all of the [\ion{C}{ii}] 158\,$\mu$m emission has clear molecular gas counterparts. Some of the extended [\ion{C}{ii}] 158\,$\mu$m emission at velocities close to the main molecular gas velocities may be tracing molecular cloud assembly, as suggested in observations \citep{SchneiderBonne:2023na}, and simulations \citep{ClarkGlover:2019aa}. Finding the line width of the [\ion{C}{ii}] 158\,$\mu$m emission component to be $\Delta\nu=3.4$\,\kms\ in G31.388$-$0.383 -- broader than the main molecular component -- may itself be a signature of molecular clouds in formation \citep{HeyerGoldsmith:2022xg}. The feature is slightly stronger ($T_{\rm mb, peak} = 0.55$\,K) than previous detections of [\ion{C}{ii}] 158\,$\mu$m emission in the diffuse ISM \citep[][]{GoldsmithPineda:2018aa}. Some of the population of features with weak [\ion{C}{ii}] 158\,$\mu$m emission may have hence been missed at our sensitivity of $\sigma\sim0.1$\,K, as well as due to line-of-sight confusion in velocity. 

The OH observations towards G29.935$-$0.053 appear to resemble parts of the common phenomenon ``satellite line reversal'' seen in several \ion{H}{ii} regions \citep{HafnerDawson:2020qw}, which is an opposing pattern of the OH satellite lines where the OH 1612 MHz line changes from absorption to inversion with increasing velocity, while the OH 1720 MHz changes from inversion to absorption. Even though not all parts of the profile are significantly detected, the identified parts of the features are in agreement with this phenomenon. Also, incomplete detections of the full profile in the OH satellite lines appear to be common \citep{RugelBeuther:2018aa}. The resemblance with the global pattern makes the observed line profiles in G29.935$-$0.053 unlikely to be related to any geometrical alignments unique to this object. \citet{HafnerDawson:2020qw} attribute the OH 1612 MHz inversion layer to a quiescent gas component into which the \ion{H}{ii} region is expanding, while the blueshifted OH 1720 MHz inversion originates from shocked gas itself.  A detailed account of such an expansion scenario has been given in M17, as presented in \citet[][see also \citealt{BroganTroland:2001kf,BroganTroland:1999yt}]{PellegriniBaldwin:2007aa}, who find blueshifted OH and \ion{H}{i} absorption tracing an expanding shell into the ambient ISM. The ambient gas is associated with a redshifted component. Spatially, M17 also shows arcs in \ion{H}{i} and OH absorption associated with the expansion. Applied to G29.935$-$0.053, the extended redshifted component would represent the molecular medium the \ion{H}{ii} region is expanding into, while the blueshifted absorption would be associated with the expanding shell itself. Albeit no proper 3D model of the \ion{H}{ii} complex is available for a full confirmation, the scenario proposed by \citet{HafnerDawson:2020qw} seems to be largely in agreement with the observations found here. A potential caveat would be that the blueshifted OH absorption component appears to coincide with the velocity of the dense gas clump towards pixel 5, for which the blueshifted component could be associated with the dense molecular gas itself which the HII region is expanding into at one side, instead being solely related to the expansion of the shell. A larger sample of resolved studies of OH absorption can shed further light on this problem.  

\section{Conclusion} \label{sec:conclusion}
We present high-sensitivity OH and \ion{H}{i} absorption observations at cm-wavelengths with the VLA towards four lines of sight. These are follow-up observations of the THOR survey, initially to investigate faint OH absorption from diffuse interstellar clouds and cloud envelopes, but which were found to also trace OH in high-mass star forming regions. For two lines of sight, we compared the OH lines with [\ion{C}{ii}] 158\,$\mu$m emission observations from upGREAT at SOFIA, as well as deep CO ($2-1$) lines with the PI230 receiver at the APEX telescope. 
\begin{itemize}
\item
We find OH absorption at 1665 MHz and 1667 MHz in a variety of ISM environments, both associated with dense and diffuse gas, and with gas with intermediate- and high- molecular gas fractions. We detect one feature which is without a CO counterpart. We compare OH with hydrogen column densities as derived from CO and HI. Within uncertainties we find $N_{\rm OH}$ linearly correlated both with respect to $N_{\rm H}$, where this is the total hydrogen column density, with a fitted power-law exponent of $m_{\rm H}=1.2\pm{0.1}$, and with respect to $N_{\rm H_2}$ ($m_{\rm H_2}=1.0\pm{0.1}$). The derived OH abundances, X$_{\rm OH, H_2} =N_{\rm OH}/N_{\rm H_2} = 1.2^{+0.3}_{-0.2}\times10^{-7}$ and {$X_{\rm OH, H} = N_{\rm OH}/(2\times N_{\rm H_2} + N_{\rm HI}) = 4.8^{+0.9}_{-0.8}\times10^{-8}$}, agree well with previous studies within the uncertainties and scatter of our measurements.
\item
We detect [\ion{C}{ii}] 158\,$\mu$m emission at similar velocities as an OH absorption feature towards G31.388$-$0.383. The estimated column density of ionized carbon exceeds the column density of carbon monoxide, indicating that significant fractions of the carbon in this feature are ionized, although the fraction of C$^{+}$ gas associated with the OH absorption is not clear. Other associations with line-of-sight OH absorption are not detected or remain ambiguous. G29.935$-$0.053 shows [\ion{C}{ii}] 158\,$\mu$m emission from PDRs in the \ion{H}{ii} region complex with indications of [\ion{C}{ii}] 158\,$\mu$m self-absorption, which is potentially associated with OH absorption. 
\item
G29.935$-$0.053 shows spectrally varying excitation conditions in all four OH lines, which manifest themselves in different spatial morphologies of the optical depths of the OH satellite lines (OH 1612/1720 MHz) as compared to the main lines (OH 1665/1667 MHz). This confirms previous studies in that both OH  absorption and stimulated emission can originate from multiple gas components in an \ion{H}{ii} region complex. 
\end{itemize}
For the mid-plane of the Milky Way disk, we find that the detection of true CO-dark gas remains challenging. The relevance of the hyperfine splitting transitions of hydroxyl at 18\,cm as tracer of the onset of molecular cloud formation is yet to be further explored. Ray-tracing of simulations, observations of the OH 18\,cm lines with future radio telescopes, or complementary investigations of transitions of OH at other frequencies (e.g., \citealt{JacobNeufeld:2022aa}, Busch et al., in prep.) will help to address this question. 

\bibliographystyle{aa}
\bibliography{oh_highres_paper.bib}

\begin{acknowledgements}
We dedicate this work to Karl M. Menten, who significantly contributed to the present manuscript and sadly passed away before the submission of the paper. He was a brilliant and accomplished scientist, a mentor of the highest standard, and will be greatly missed. We would like to thank the anonymous referee for comments and suggestions which greatly improved this work. M.R.R thanks several members of the MPIfR in Bonn, the Center for Astrophysics | Harvard \& Smithsonian, and the HyGAL collaboration for discussions, in particular H. Wiesemeyer, M. Reid, Q. Zhang, P. Myers, A. Jacob, M. Busch, W. Kim, D. Neufeld, Brian Svoboda. The authors would like to thank the SOFIA and APEX teams for carrying out the observations, and the support in preparing them. The National Radio Astronomy Observatory is a facility of the National Science Foundation operated under cooperative agreement by Associated Universities, Inc. Based in part on observations made with the NASA/DLR Stratospheric Observatory for Infrared Astronomy (SOFIA). SOFIA is jointly operated by the Universities Space Research Association, Inc. (USRA), under NASA contract NNA17BF53C , and the Deutsches SOFIA Institut (DSI) under DLR contract 50 OK 2002 to the University of Stuttgart. The development and operation of GREAT was financed by resources from the participating institutes, by the Deutsche Forschungsgemeinschaft (DFG) within the grant for the Collaborative Research Center 956 as well as by the Federal Ministry of Economics and Energy (BMWI) via the German Space Agency (DLR) under Grants 50 OK 1102, 50 OK 1103 and 50 OK 1104. This research was conducted in part at the Jet Propulsion Laboratory, which is operated by the California Institute of Technology under contract with the National Aeronautics and Space Administration (NASA).
This data made use of splatalogue, CDS, simbad, astropy, reproject, spectral-cube, aplpy, matplotlib, gildas, carta. 
\end{acknowledgements}

\clearpage

\begin{appendix}
\section{Additional spectra and results} \label{sec:spectra}

\begin{sidewaystable*}
\tiny
\centering
\caption{OH, HI, and CO column densities}
\label{tbl:oh_integrals}
\setlength{\tabcolsep}{3pt}
\begin{tabular}{lrr|rr|rr|rr|rr|rr|rr|rrrr|r}
\hline \hline
  &   & &  \multicolumn{2}{c|}{OH 1667 MHz}&  \multicolumn{2}{c|}{OH 1665 MHz}  & \multicolumn{2}{c|}{OH 1612 MHz} & \multicolumn{2}{c|}{OH 1720 MHz}& \multicolumn{2}{c|}{HI}&  & & $^{12}$CO & $^{13}$CO & $C^{18}$O & \multicolumn{1}{c|}{$^{13}$CO}& \\
  &   & &  \multicolumn{2}{c|}{}&  \multicolumn{2}{c|}{}  & \multicolumn{2}{c|}{} & \multicolumn{2}{c|}{}& \multicolumn{2}{c|}{}&  & & $J$=1--0 &$J$=1--0 &$J$=1--0 & \multicolumn{1}{c|}{$J$=1--0}&  \\
Source & $\varv_{\rm mid}$ & $R_{\rm GC}$ &  $\int\tau{\rm d}\varv$ & $\tau_{\rm max}$ & $\int\tau{\rm d}\varv$ & $\tau_{\rm max}$ & $\int\tau{\rm d}\varv$ & $\tau_{\rm max}$ & $\int\tau{\rm d}\varv$ & $\tau_{\rm max}$ & $\int\tau{\rm d}\varv$ & $\tau_{\rm max}$ &$\frac{N_{\rm OH}}{T_{\rm ex}}$ &$\frac{N_{\rm \ion{H}{i}}}{T_{\rm S}}$ & $\int T_{\rm mb} {\rm d}\varv$& $\int T_{\rm mb} {\rm d}\varv$& $\int T_{\rm mb} {\rm d}\varv$& $\int T_{\rm mb} {\rm d}\varv$& $N_{\rm H_2}$ \\
&[\kms] & [kpc]&[\kms]&&[\kms]&&[\kms]&&[\kms]&&[\kms]&& x$10^{13}\frac{{\rm cm}^{-2}}{{\rm K}}$ & x$10^{19}\frac{{\rm cm}^{-2}}{{\rm K}}$ & [K \kms] & [K \kms] & [K \kms] & [K \kms] & x$10^{20}$cm$^{-2}$ \\                                 
\hline
G21.347$-$0.629  
& +7.4 & 7.64 & +0.25 & +0.14 & +0.14 & +0.07 & $<$0.07 & $<$0.01 & +0.03 & +0.02 & $>$18.95 & $>$3.66 &  &  & 24.5 & 1.5 & $<$1.3 & 1.5 & 49.0 \\
& +51.6 & 5.07 & +0.36 & +0.14 & +0.17 & +0.07 & +0.00 & +0.01 & +0.06 & +0.04 & $>$10.82 & $>$3.64 & 8.09 & $>$1.97 & 51.9 & 11.5 & $<$1.3 & 11.5 & 103.7 \\
& +56.5 & 4.87 & +0.50 & +0.37 & +0.32 & +0.28 & -0.04 & +0.00 & +0.15 & +0.13 & $>$9.04 & $>$3.70 & 11.17 & $>$1.65 & 22.2 & 9.7 & 2.9 & 9.7 & 44.5 \\
& +73.5 & 4.24 & +0.13 & +0.04 & +0.06 & +0.03 & +0.04 & +0.02 & $<$0.06 & $<$0.01 & +7.31 & +1.38 & 2.86 & 1.33 & 16.1 & $<$1.3 & $<$1.3 & $<$1.3 & 32.1 \\
& +68.8 & 4.41 & +0.05 & +0.06 & +0.02 & +0.03 & $<$0.02 & $<$0.01 & $<$0.02 & $<$0.01 & +0.58 & +0.36 & 1.13 & 0.11 & $<$2.4 & $<$1.3 & $<$1.3 & $<$1.3 & $<$4.9 \\
& +66.6 & 4.48 & +0.05 & +0.05 & +0.03 & +0.03 & $<$0.02 & $<$0.01 & $<$0.02 & $<$0.01 & +0.50 & +0.27 & 1.07 & 0.09 & 2.2 & $<$1.3 & $<$1.3 & $<$1.3 & 4.4 \\
\hline
G29.957$-$0.018 
& +6.5 & 7.87 & +0.19 & +0.12 & +0.11 & +0.07 & +0.08 & +0.06 & -0.03 & +0.01 & +4.19 & +1.56 & 4.27 & 0.76 & 12.9 & 2.2 & $<$1.3 & 2.2 & 25.8 \\
& +9.7 & 7.68 & +0.08 & +0.07 & +0.05 & +0.04 & +0.04 & +0.03 & $<$0.02 & $<$0.01 & +2.67 & +1.14 & 1.84 & 0.49 & 2.6 & $<$1.3 & $<$1.3 & $<$1.3 & 5.1 \\
& +67.5 & 5.18 & +0.10 & +0.03 & +0.05 & +0.02 & +0.04 & +0.02 & $<$0.06 & $<$0.01 & +7.55 & +1.26 & 2.15 & 1.38 & 21.9 & 4.3 & $<$1.3 & 4.3 & 43.8 \\
& +97.8 & 4.32 & +0.14 & +0.06 & +0.09 & +0.04 & -0.03 & +0.02 & +0.08 & +0.06 & $>$13.24 & $>$3.59 & 3.18 & $>$2.41 & 160.6 & 63.9 & 13.9 & 63.9 & 321.1$^{*}$\\
& +102.8 & 4.19 & +0.14 & +0.06 & -0.25 & +0.01 & -0.19 & +0.00 & +0.16 & +0.07 & $>$12.18 & $>$3.59 & 3.21 & $>$2.22 & 87.1 & 24.6 & 4.4 & 24.6 & 174.1$^{*}$\\
\hline
\hline
  &&   &  \multicolumn{2}{c|}{}&  \multicolumn{2}{c|}{}  & \multicolumn{2}{c|}{} & \multicolumn{2}{c|}{}& \multicolumn{2}{c|}{}&  & & $^{12}$CO & $^{13}$CO & $C^{18}$O & \multicolumn{1}{c|}{}&  \\
  &&   &  \multicolumn{2}{c|}{}&  \multicolumn{2}{c|}{}  & \multicolumn{2}{c|}{} & \multicolumn{2}{c|}{}& \multicolumn{2}{c|}{}&  & & $J$=2--1 &$J$=2--1 &$J$=2--1 & \multicolumn{1}{c|}{}& \\
\hline
G29.935$-$0.053 
& +7.5 & 7.81 & +0.15 & +0.07 & +0.09 & +0.05 & +0.08 & +0.05 & $<$0.12 & $<$0.04 & +5.53 & +1.29 & 3.47 & 1.01 & 3.4 & 0.4 & $<$0.2 & $<$1.3 & 9.8 \\
& +50.8 & 5.77 & +0.29 & +0.14 & +0.13 & +0.06 & $<$0.12 & $<$0.04 & $<$0.11 & $<$0.04 & $>$11.84 & $>$2.64 & 6.39 & $>$2.16 & 22.5 & 5.6 & 0.8 & 7.8 & 64.3 \\
& +68.0 & 5.17 & +0.20 & +0.14 & +0.13 & +0.09 & +0.09 & +0.05 & $<$0.08 & $<$0.04 & $>$8.30 & $>$2.62 & 4.56 & $>$1.51 & 9.9 & 1.1 & $<$0.2 & 3.2 & 28.4 \\
& +99.0 & 4.29 & +0.76 & +0.26 & +0.64 & +0.18 & -0.07 & +0.03 & +0.27 & +0.08 & $>$18.10 & $>$2.64 & 17.07 & $>$3.30 & 113.7 & 58.1 & 18.2 & 72.3 & 377.5$^{a,*}$ \\
\hline
G31.388$-$0.383  
& -3.8 & 8.54 & +0.10 & +0.07 & +0.04 & +0.03 & +0.03 & +0.02 & $<$0.03 & $<$0.02 & +1.16 & +0.57 & 2.27 & 0.21 & 1.5 & 0.2 & $<$0.2 & $<$0.9 & 4.3 \\
& +18.0 & 7.25 & +0.22 & +0.14 & +0.11 & +0.08 & +0.04 & +0.03 & +0.02 & +0.02 & $>$22.74 & $>$3.89 & 4.98 & $>$4.14 & 7.6 & 0.6 & $<$0.2 & $<$0.9 & 21.8 \\
& +37.0 & 6.38 & +0.03 & +0.04 & $<$0.03 & $<$0.02 & $<$0.03 & $<$0.02 & $<$0.03 & $<$0.02 & +2.15 & +0.88 & 0.74 & 0.39 & 0.3 & $<$0.2 & $<$0.2 & $<$0.9 & 0.8 \\
& +73.1 & 5.11 & +0.12 & +0.04 & +0.06 & +0.03 & +0.07 & +0.02 & -0.06 & +0.00 & +17.18 & +3.59 & 2.58 & 3.13 & 7.3 & 0.5 & $<$0.2 & 2.5 & 20.9 \\
& +78.1 & 4.96 & +0.12 & +0.12 & +0.07 & +0.07 & +0.03 & +0.02 & +0.01 & +0.02 & +5.67 & +2.39 & 2.64 & 1.03 & 2.0 & 0.2 & $<$0.2 & 0.7 & 5.8 \\
& +81.8 & 4.85 & +0.06 & +0.04 & $<$0.03 & $<$0.02 & $<$0.03 & $<$0.02 & $<$0.03 & $<$0.02 & $>$11.84 & $>$3.85 & 1.37 & $>$2.16 & 3.6 & 0.4 & $<$0.2 & 1.0 & 10.4 \\
& +85.3 & 4.75 & +0.11 & +0.07 & +0.03 & +0.03 & $<$0.03 & $<$0.02 & +0.03 & +0.02 & $>$13.04 & $>$3.87 & 2.53 & $>$2.38 & 2.8 & 0.2 & $<$0.2 & 0.4 & 7.9 \\
& +92.0 & 4.57 & +0.05 & +0.04 & +0.04 & +0.02 & $<$0.03 & $<$0.01 & $<$0.03 & $<$0.02 & +6.65 & +2.47 & 1.19 & 1.21 & 4.7 & 0.5 & $<$0.2 & 1.5 & 13.4 \\
\hline
\end{tabular}
\tablefoot{Integral and peak of the optical depth of the OH and \ion{H}{I} lines, as well as of CO emission in specific velocity intervals with the mid-point of each chosen velocity range tabulated in column 1. The CO isotopologue and transition reported here and used to determine the H$_2$ column density is indicated in brackets in the last column. As we do not derive the excitation temperatures of the OH and \ion{H}{I} lines in this work, we report $N_{\rm OH}/T_{\rm ex}$ and $N_{\rm HI}/T_{\rm S}$. $(^{*})$ Features associated with \ion{H}{ii} regions. All other features are line-of-sight absorption features. $({}^a)$ Using $^{13}$CO $J=1-0$ and correcting for isotope ratio variations over Galactocentric distance (Sect.~\ref{sec:analysis:co}).}
\end{sidewaystable*}

\begin{figure*}[!ht]
\centering
  \includegraphics[width=0.99\textwidth, trim=0 3cm 0 0]{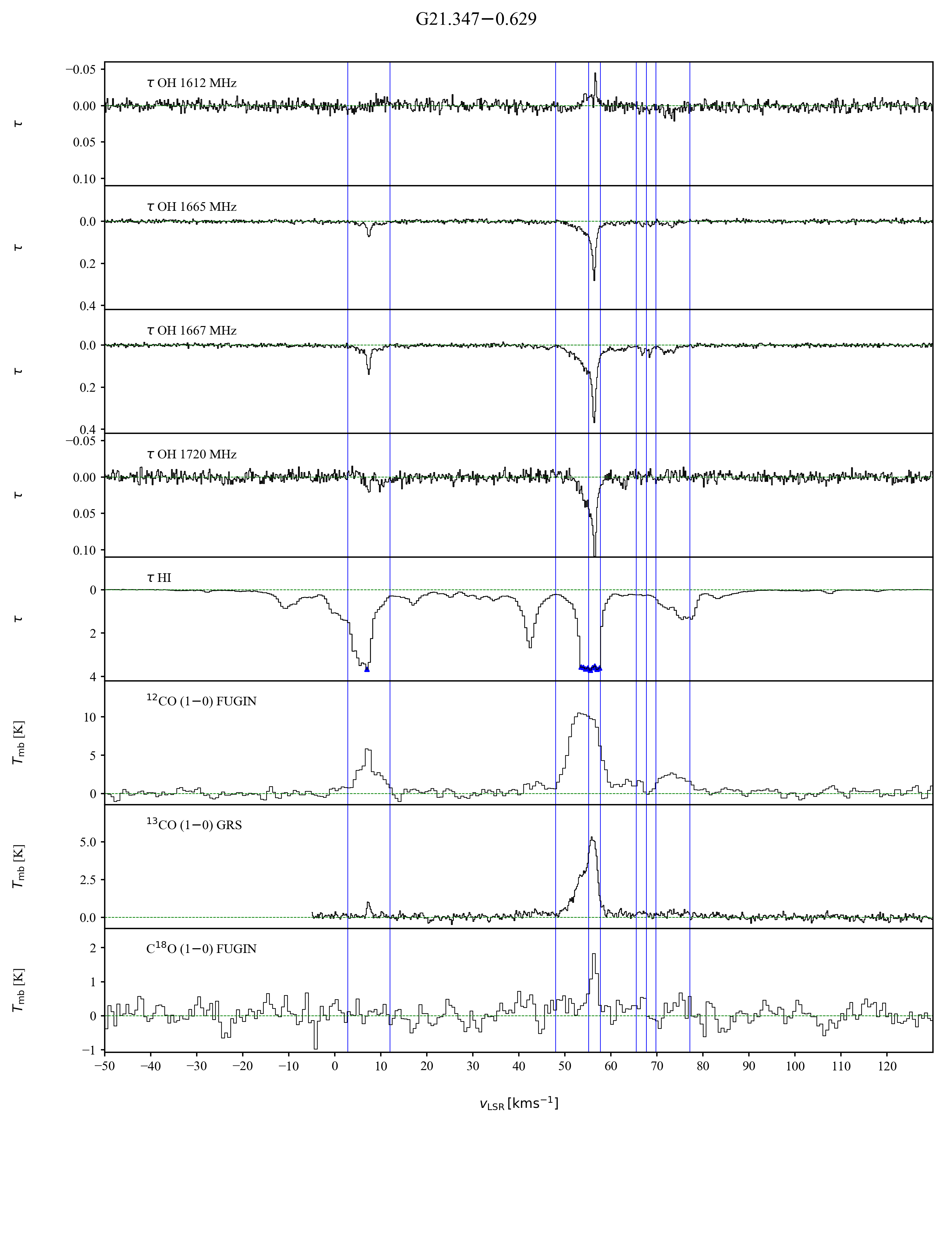}
\caption{
Same as Fig.~\ref{fig:overview_transitions_g31}, but towards G21.347$-$0.629, except the CO data presented is taken from FUGIN \citep{UmemotoMinamidani:2017aa} and GRS \citep{JacksonRathborne:2006aa} surveys. No [\ion{C}{ii}] 158\,$\mu$m emission spectrum has been recorded for this source.
}
  \label{fig:overview_transitions_g21}
\end{figure*}

\begin{figure*}[!ht]
\centering
  \includegraphics[width=0.99\textwidth, trim=0 3cm 0 0]{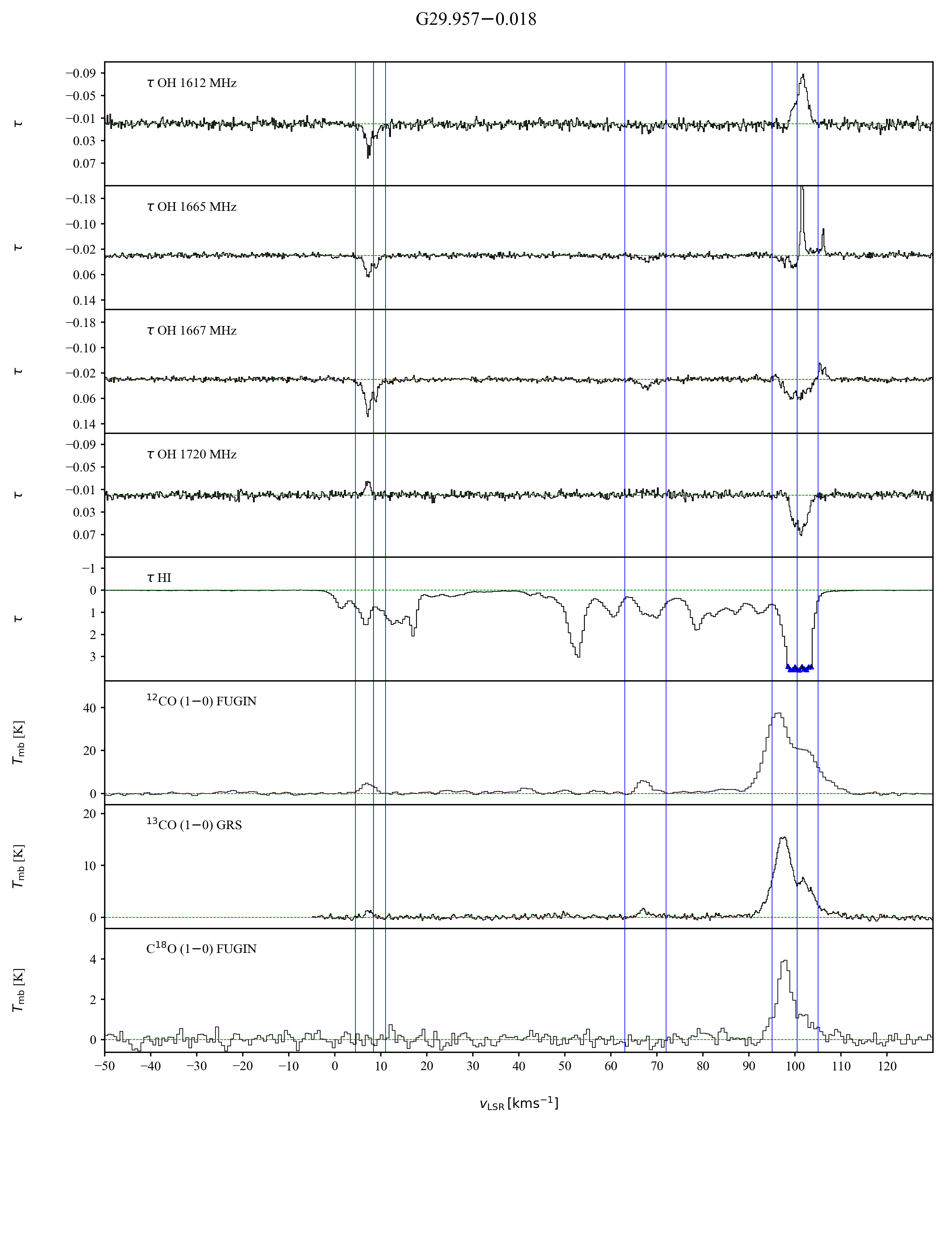}
\caption{Same as Fig.~\ref{fig:overview_transitions_g21}, but for G29.957$-$0.018}
  \label{fig:overview_transitions_g29}
\end{figure*}

\begin{figure*}[!ht]
\centering
  \includegraphics[width=0.99\textwidth, trim=0 0 0 0]{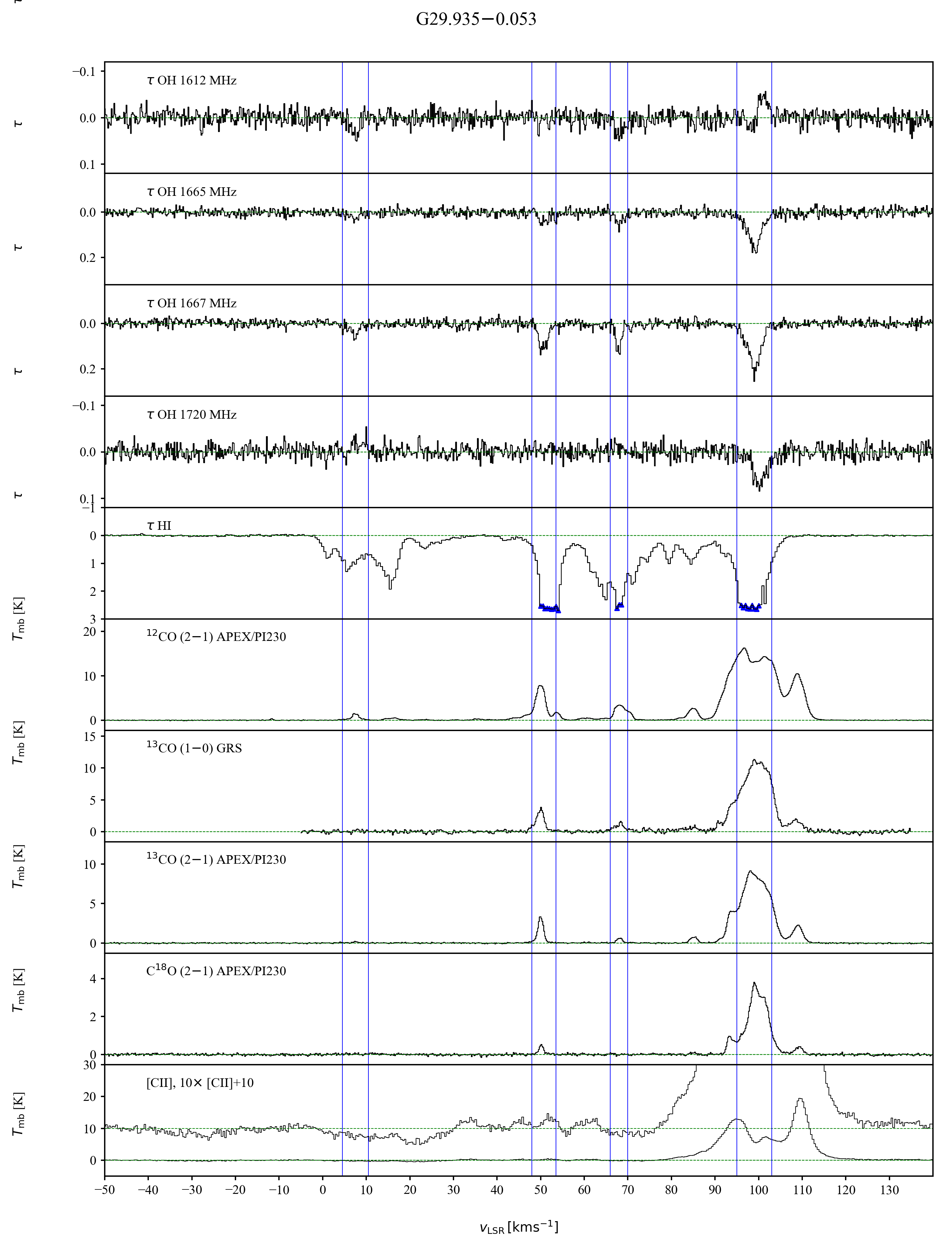}
\caption{Same as Fig.~\ref{fig:overview_transitions_g31}, but for G29.935$-$0.053. The bottom panel includes an enlarged view of the [\ion{C}{ii}] 158\,$\mu$m line for better visibility of the faint emission or absorption components, and/or baseline ripples.}
  \label{fig:overview_transitions_g29p935}
\end{figure*}

\begin{figure}
\centering
\includegraphics[width=1\columnwidth]{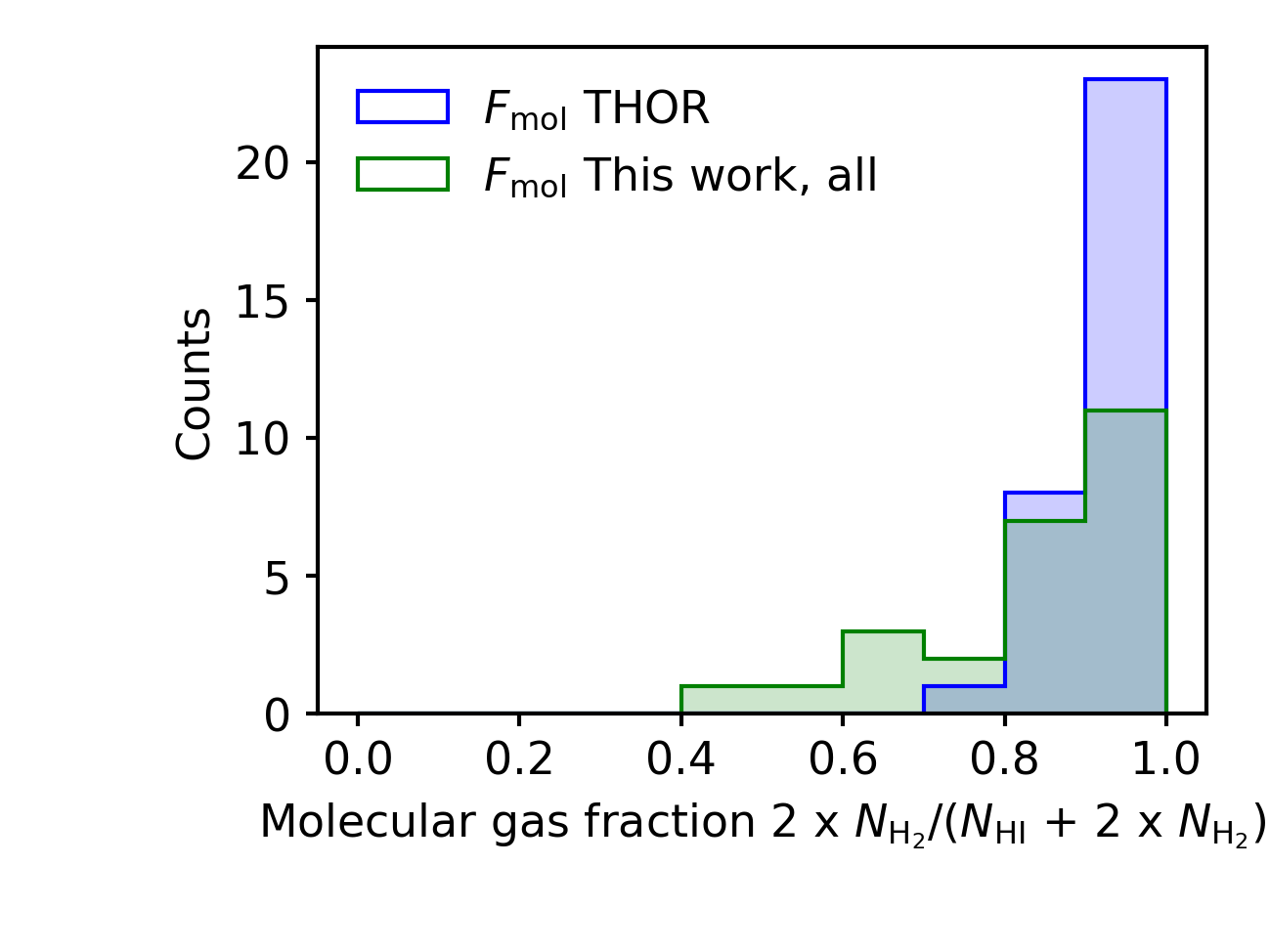}
\caption{As Fig.~\ref{fig:fmol_thorhr_thor}, but for $T_{\rm spin} = 50\,{\rm K}$.}
\label{fig:fmol_thorhr_thor_50k}
\end{figure}

\begin{figure*}[!ht]
\centering
  \includegraphics[height=1.1\textwidth]{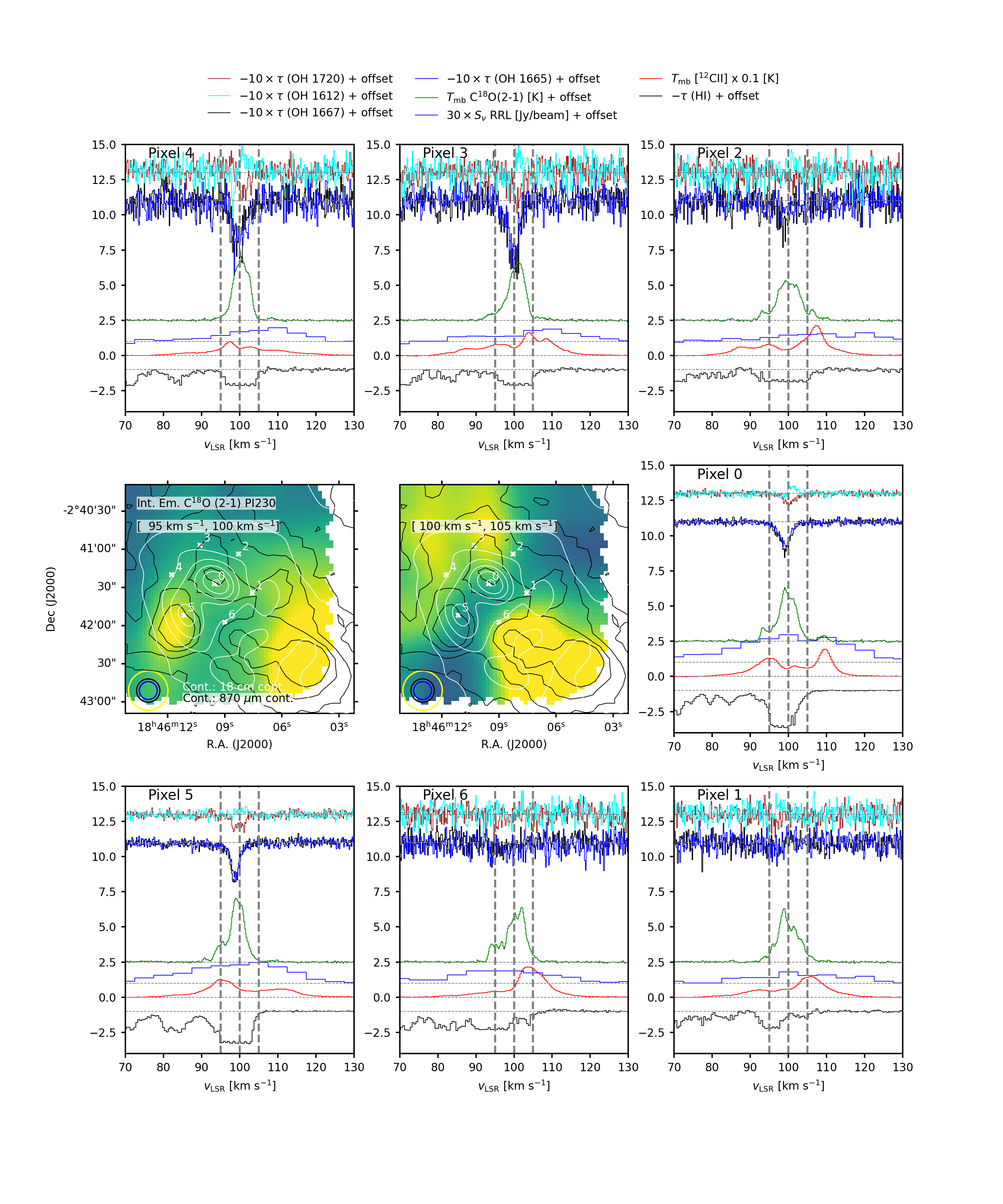}
\caption{Optical depth of the OH lines at 1612 MHz, 1665 MHz, 1667 MHz and 1720 MHz, as well of \ion{H}{i} towards the seven pixels towards G29.935$-$0.053 of the upGREAT/SOFIA [$^{12}$\ion{C}{ii}] observations, as well as accompanying C$^{18}$O (2--1) (APEX/PI230) and stacked radio recombination lines (40$^{\prime\prime}$, GLOSTAR; \citealt{BrunthalerMenten:2021aa,KhanRugel:2024ju}). The maps in the middle panel are repeated for clarity from row 2 in Fig.~\ref{fig:mommap_g29_all}.}
  \label{fig:g29_spec}
\end{figure*}

\section{Overview SOFIA data} \label{sec:app_sofia}

\begin{figure*}[!ht]
\centering
  \includegraphics[width=0.9\textwidth]{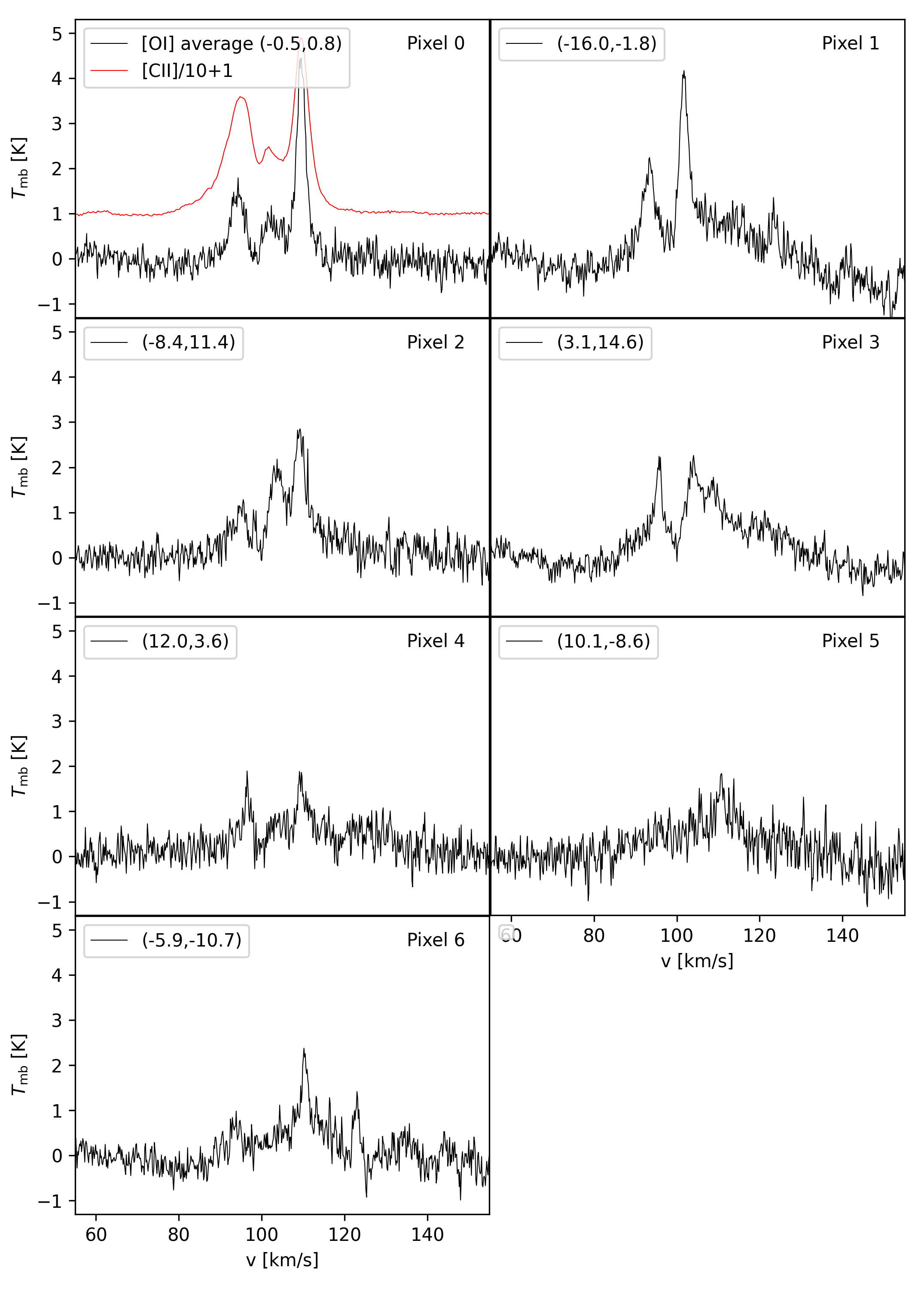}
\caption{[\ion{O}{i}] 63\,$\mu$m emission towards G29.935$-$0.053. For each spectrum, we show the average of all data ({\it black}). We only show the [\ion{C}{ii}] 158\,$\mu$m emission for reference in pixel 1, as the other pixels are closer to the central beam than for [\ion{C}{ii}] 158\,$\mu$m and do not overlap. Offsets from the source position are indicated in arcsec.}
  \label{fig:g29p935_oi}
\end{figure*}

\begin{figure*}[!ht]
\centering
  \includegraphics[width=0.9\textwidth]{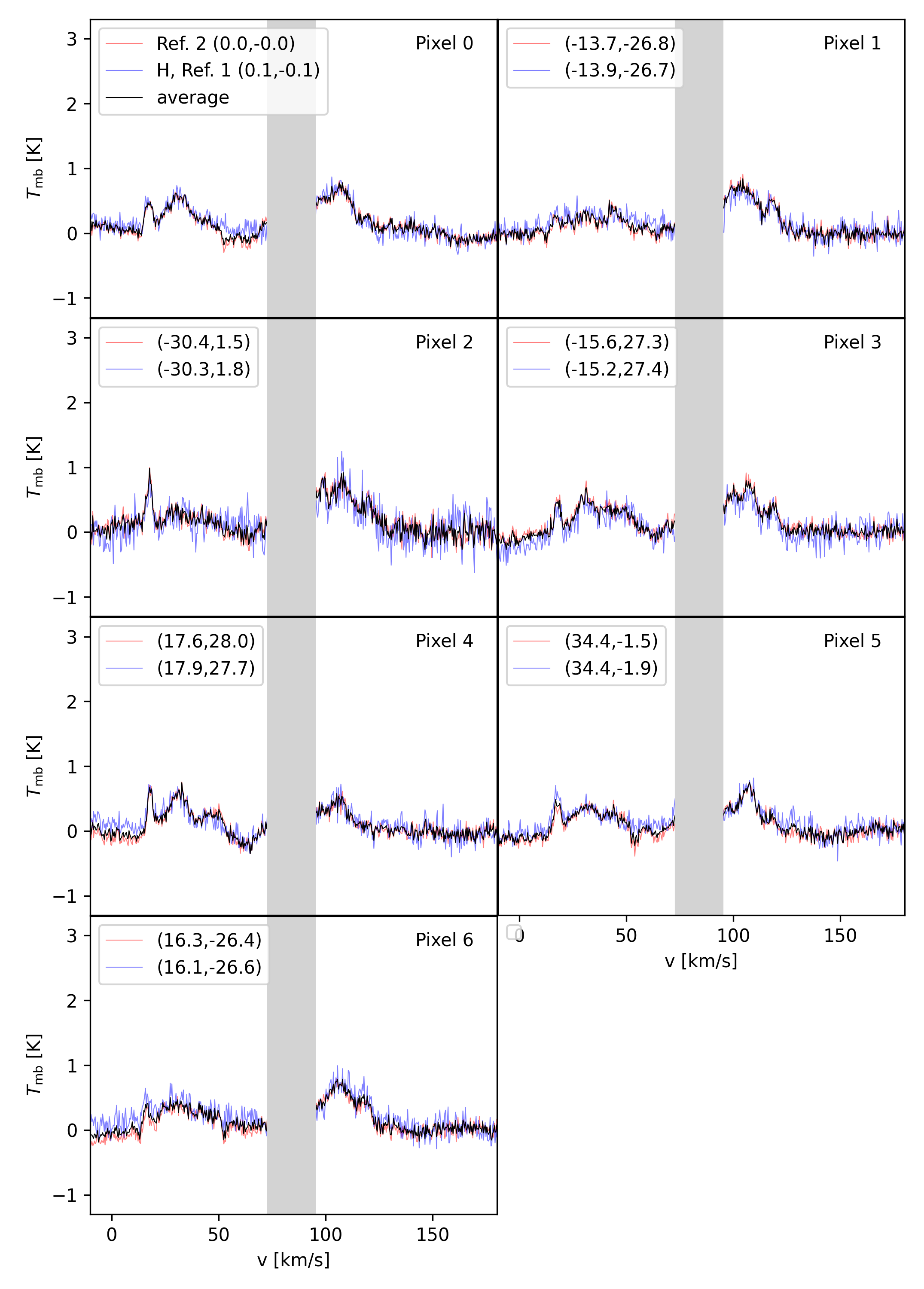}
\caption{[\ion{C}{ii}] 158\,$\mu$m emission towards G31.388$-$0.383. For each spectrum, we show the average of all data ({\it black}), as well as averages by polarization and by reference position. Note that only vertical polarization (``V'') is available for Pixel 2. The gray-shaded region is masked as it is contaminated by emission in the reference position.}
  \label{fig:g31p388_cii}
\end{figure*}

\begin{figure*}[!ht]
\centering
  \includegraphics[width=0.9\textwidth]{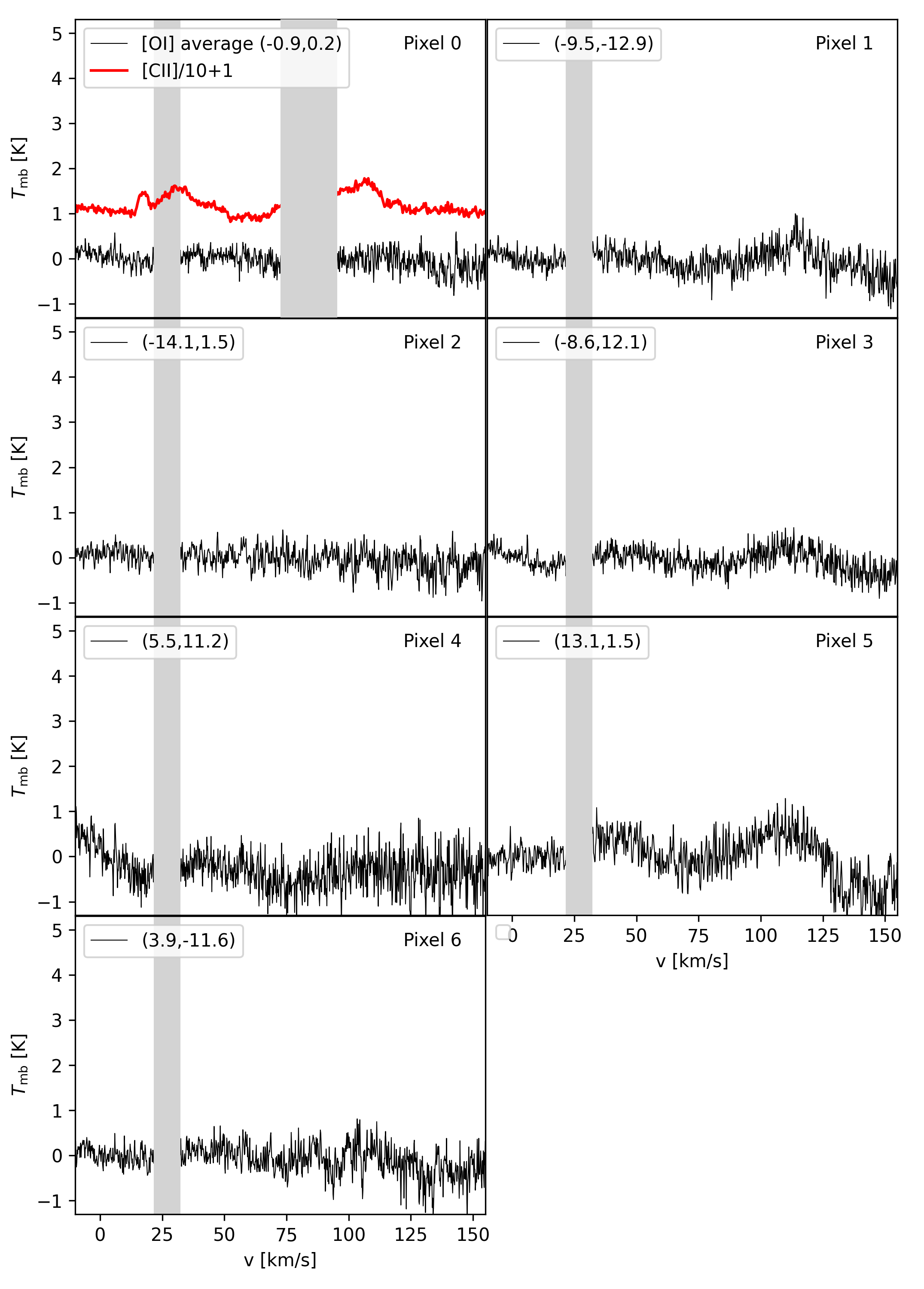}
\caption{[\ion{O}{i}] 63\,$\mu$m emission towards G31.388$-$0.383. For each spectrum, we show the average of all data ({\it black}). We only show the [\ion{C}{ii}] 158\,$\mu$m emission for reference in pixel 1, as the other pixels are closer to the central beam than for [\ion{C}{ii}] 158\,$\mu$m and do not overlap.  Several pixels show residual low-frequency baseline ripples. The gray-shaded region around 25 \kms\ is masked due to residual artefacts from atmospheric corrections.}
  \label{fig:g31p388_oi}
\end{figure*}

\begin{figure*}[!ht]
\centering
  \includegraphics[width=0.9\textwidth]{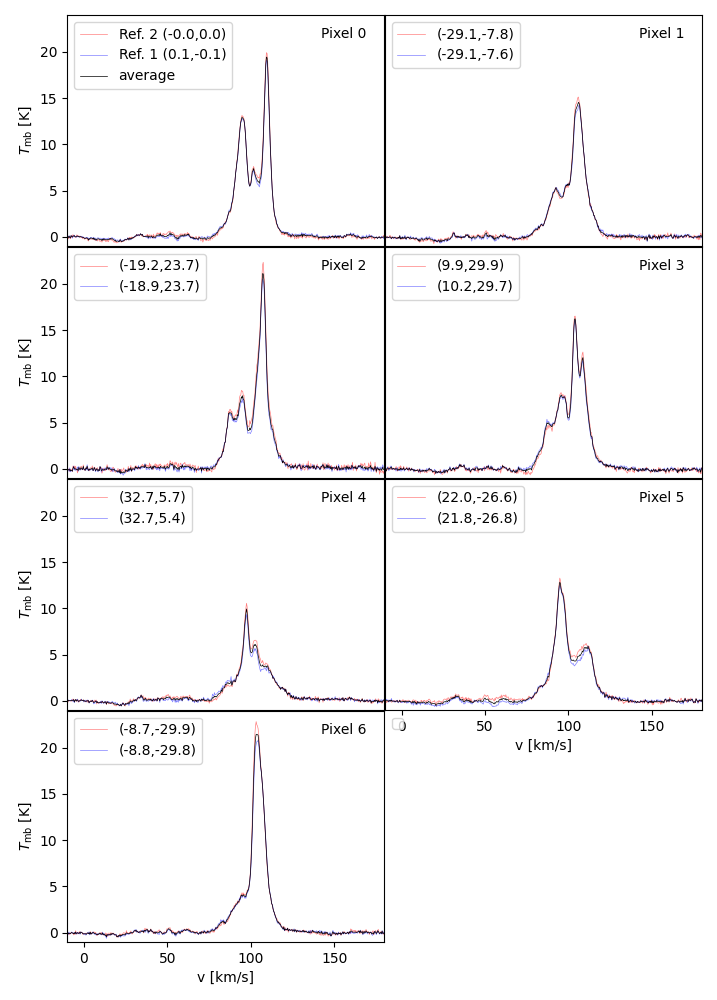}
\caption{[\ion{C}{ii}] 158\,$\mu$m emission towards G29.935$-$0.053. For each spectrum, we show the average of all data ({\it black}), as well as averages by reference position. Note that only vertical polarization (``V'') is available for Pixel 2. The pixel offset from the "H"The gray-shaded region is masked as it is contaminated by emission in the reference position.}
  \label{fig:g29p935_cii}
\end{figure*}

\begin{figure*}[!ht]
\centering
  \includegraphics[width=0.9\textwidth]{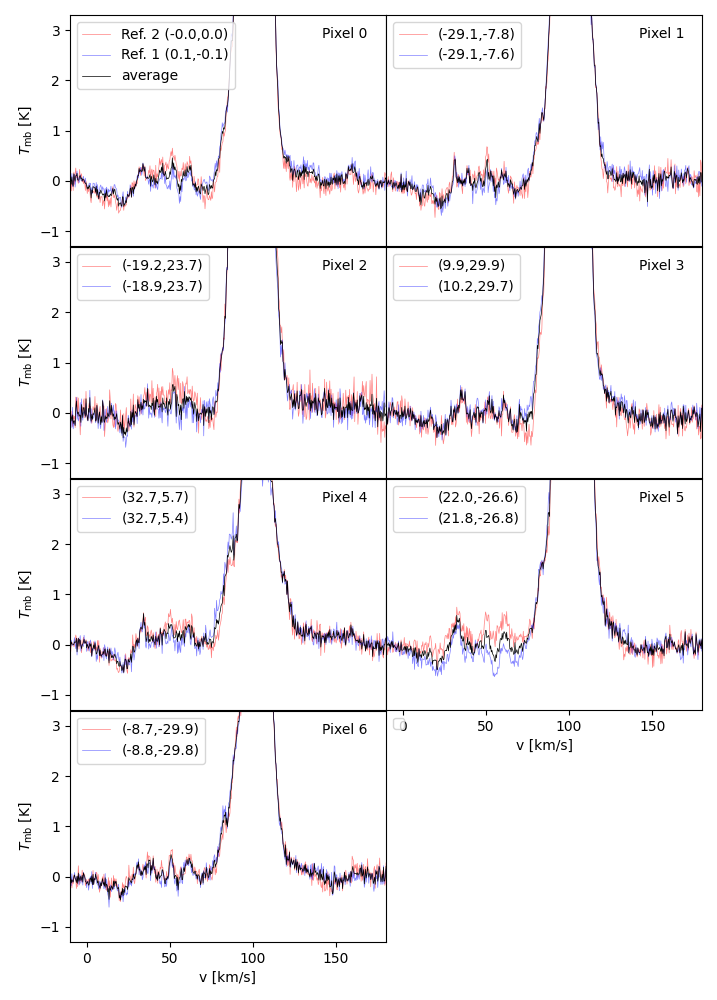}
\caption{Zoom-in on the low-level [\ion{C}{ii}] 158\,$\mu$m emission in G29.935$-$0.053. Colors and spectra as in Fig.~\ref{fig:g29p935_cii}.}
  \label{fig:g29p935_cii_zoomlow}
\end{figure*}

\begin{figure*}[!ht]
\centering
  \includegraphics[width=0.9\textwidth]{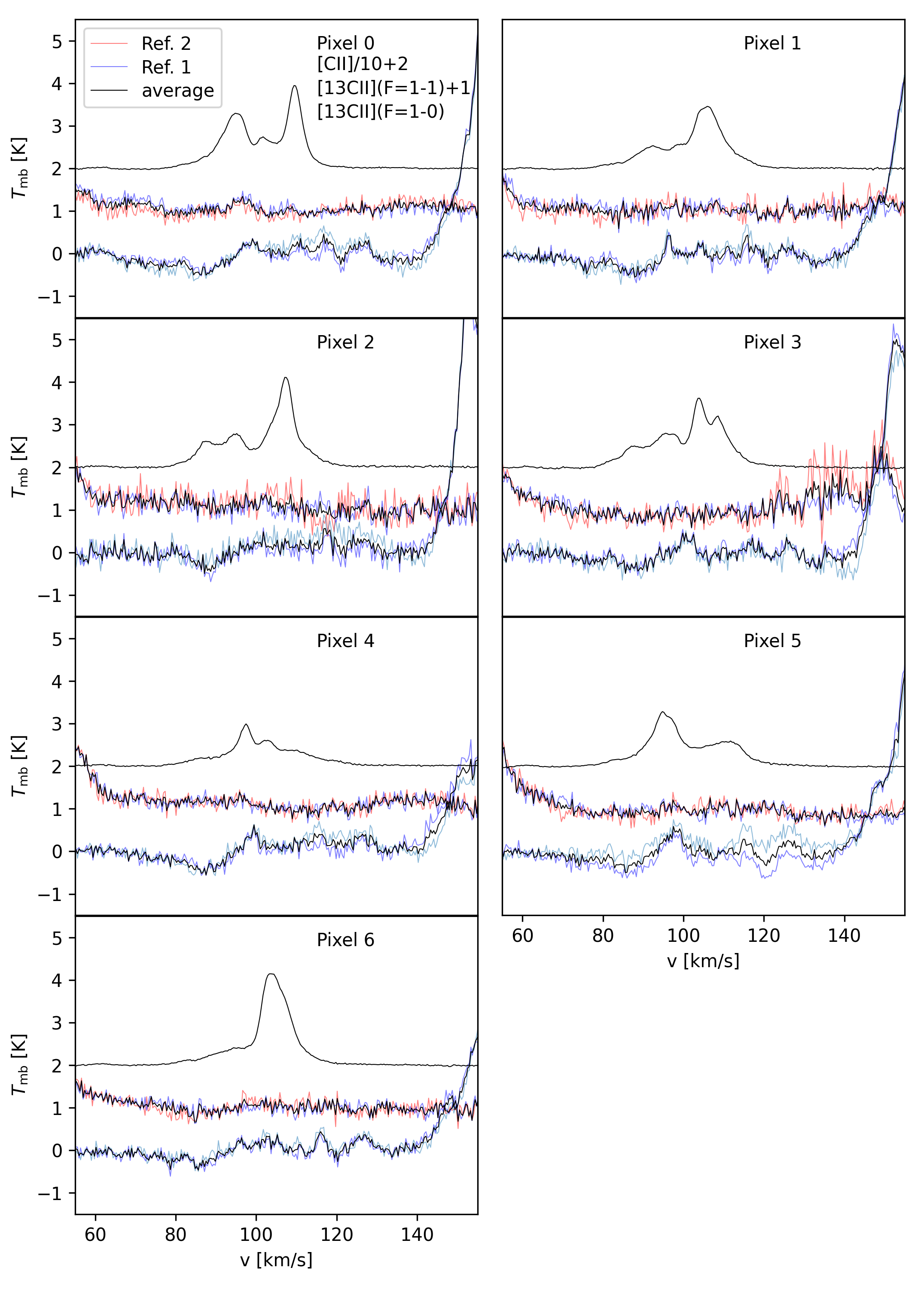}
\caption{[\ion{C}{ii}] 158\,$\mu$m and [${}^{13}$\ion{C}{ii}] emission towards G29.935$-$0.053. Colors and spectra as in Fig.~\ref{fig:g29p935_cii}, with the spectra in each subplot shifted to the rest frequency of each line.}
  \label{fig:g29p935_cii_13cii}
\end{figure*}
\end{appendix}

\end{document}